\documentclass[preprint2]{aastex63}
\usepackage{etoolbox}
\usepackage{graphicx}
\usepackage{natbib}

\newcommand{\gsim}{\mbox{\hspace{.2em}\raisebox{.5ex}{$>$}\hspace{-.8em}\raisebox{-.5ex}{$\sim$}\hspace{.2em}}}
\newcommand{\lsim}{\mbox{\hspace{.2em}\raisebox{.5ex}{$<$}\hspace{-.8em}\raisebox{-.5ex}{$\sim$}\hspace{.2em}}}

\newcommand{\E}[1]{\times 10^{#1}}
\newcommand{\twCO}{$^{12}$CO}  \newcommand{\thCO}{$^{13}$CO}

\newcommand{\HI}{\mbox{H\,\textsc{i}}}

\newcommand{\s}{\,{\rm s}}      \newcommand{\ps}{\,{\rm s}^{-1}}
\newcommand{\yr}{\,{\rm yr}}    \newcommand{\Msun}{M_{\odot}}   
    \newcommand{\km}{\,{\rm km}}
\newcommand{\kpc}{\,{\rm kpc}}
        \newcommand{\K}{\,{\rm K}}
    
    \newcommand{\G}{\,{\rm G}}

\begin{document}

\title{
Gas Transfer Between the Inner 3 kpc Disk and the Galactic Central Molecular Zone
}

\shorttitle{Gas Flows toward the CMZ}

\correspondingauthor{Yang Su}
\email{yangsu@pmo.ac.cn}

\author[0000-0002-0197-470X]{Yang Su}
\affil{Purple Mountain Observatory, Chinese Academy of Sciences, 
10 Yuanhua Road, Nanjing 210023, China}
\affiliation{School of Astronomy and Space Science, University of Science and
Technology of China, 96 Jinzhai Road, Hefei 230026, China}

\author{Shiyu Zhang}
\affiliation{Purple Mountain Observatory, Chinese Academy of Sciences, 
10 Yuanhua Road, Nanjing 210023, China}
\affiliation{School of Astronomy and Space Science, University of Science and
Technology of China, 96 Jinzhai Road, Hefei 230026, China}

\author{Yan Sun}
\affiliation{Purple Mountain Observatory, Chinese Academy of Sciences, 
10 Yuanhua Road, Nanjing 210023, China}
\affiliation{School of Astronomy and Space Science, University of Science and
Technology of China, 96 Jinzhai Road, Hefei 230026, China}

\author{Ji Yang}
\affiliation{Purple Mountain Observatory, Chinese Academy of Sciences, 
10 Yuanhua Road, Nanjing 210023, China}
\affiliation{School of Astronomy and Space Science, University of Science and
Technology of China, 96 Jinzhai Road, Hefei 230026, China}

\author{Fujun Du}
\affiliation{Purple Mountain Observatory, Chinese Academy of Sciences,
10 Yuanhua Road, Nanjing 210023, China}
\affiliation{School of Astronomy and Space Science, University of Science and
Technology of China, 96 Jinzhai Road, Hefei 230026, China}

\author{Min Fang}
\affiliation{Purple Mountain Observatory, Chinese Academy of Sciences,
10 Yuanhua Road, Nanjing 210023, China}
\affiliation{School of Astronomy and Space Science, University of Science and
Technology of China, 96 Jinzhai Road, Hefei 230026, China}

\author{Qing-Zeng Yan}
\affiliation{Purple Mountain Observatory, Chinese Academy of Sciences, 
10 Yuanhua Road, Nanjing 210023, China}

\author{Shaobo Zhang}
\affiliation{Purple Mountain Observatory, Chinese Academy of Sciences, 
10 Yuanhua Road, Nanjing 210023, China}

\author{Zhiwei Chen}
\affiliation{Purple Mountain Observatory, Chinese Academy of Sciences, 
10 Yuanhua Road, Nanjing 210023, China}

\author{Xuepeng Chen}
\affiliation{Purple Mountain Observatory, Chinese Academy of Sciences, 
10 Yuanhua Road, Nanjing 210023, China}
\affiliation{School of Astronomy and Space Science, University of Science and
Technology of China, 96 Jinzhai Road, Hefei 230026, China}

\author{Xin Zhou}
\affiliation{Purple Mountain Observatory, Chinese Academy of Sciences, 
10 Yuanhua Road, Nanjing 210023, China}

\author{Lixia Yuan}
\affiliation{Purple Mountain Observatory, Chinese Academy of Sciences, 
10 Yuanhua Road, Nanjing 210023, China}

\author{Yuehui Ma}
\affiliation{Purple Mountain Observatory, Chinese Academy of Sciences, 
10 Yuanhua Road, Nanjing 210023, China}

\begin{abstract}
We uncovered a more tilted molecular gas structure with highly negative 
velocities located near the dust lane. Our observations show 
that the approaching gas flows from the overshoot process 
are captured by the gravitational potential of the bar and then flow toward 
the Galactic central molecular zone (CMZ) through the bar channel.
The recycled gas from the overshoot effect, in conjunction with freshly 
accreted gas from the inner 3 kpc disk, accumulates significantly near 
$R_{\rm GC}\sim\frac{1}{2}R_{\rm bar}$ and $R_{\rm GC}\sim\frac{2}{3}R_{\rm bar}$ 
regions by adopting a bar length of $\sim$3.2--3.4~kpc. Importantly, 
within these regions, there are frequent collisions and substantial 
angular momentum exchanges between gas flows with different trajectories.
In this scenario, the DISSIPATION processes arising from 
interactions between colliding flows, together with the varying torques
induced by the nonaxisymmetric bar, effectively transfer the angular momentum of 
viscous gas outward, thereby driving the  molecular gas to settle into the 
CMZ within about three orbital periods.
A long-term gas inflow with an average rate of $\gsim 1.1\ \Msun\yr^{-1}$, 
coupled with intense transient accretion events that exceed the average rate by
several times due to the overshoot effect, significantly regulates the gas 
distribution, physical properties, and dynamical evolution of the CMZ.
These new findings provide robust observational evidence for elucidating 
the intricate dynamics of molecular gas flows toward the CMZ. 
Our observations suggest that gas dynamics have a significant impact on the 
secular evolution of both the Milky Way and the extragalactic gas-rich galaxies.
\end{abstract}


\section{Introduction}
The central molecular zone (CMZ), which may originate from the gas inflow 
driven by the Galactic bar, hosts a huge amount of molecular gas  
\citep[i.e., 2--6$\times10^{7}\ \Msun$;][]{1998A&A...331..959D,
2002A&A...384..112L,2007A&A...467..611F,2024arXiv241017334B}
within $\sim$~200--300 pc of Sgr A$^{*}$. As the closest observable 
galactic nucleus, the CMZ provides us with a wealth of critical 
information about gas cycling in large-scale Galactic dynamics,
star formation in extreme environments, and the various physical 
processes related to Galactic evolution \citep[see the 
comprehensive reviews of][]{2021NewAR..9301630B,2023ASPC..534...83H}.
Such studies on the Galactic CMZ are also of great significance for 
understanding the gas dynamics in other galaxies \citep[e.g., the case NGC 1365 in]
[]{2023ApJ...944L..15S}, as well as the evolution of galaxies and 
their own nuclei \citep[e.g.,][]
{2021ApJS..257...43L,2023A&A...676A.113S,2024ARA&A..62..369S}.

Molecular gas is the dominant gas component in the inner region of 
the Milky Way \citep[e.g.,][]{2016ApJ...823...76K}. Through spectral 
observations of molecular gas, we can investigate the distribution, 
kinematics, and other properties of molecular clouds (MCs) in detail. 
Although \twCO\ does not reveal the internal density structure of 
MCs well, it is a good indicator of the extended structure of MCs 
and a useful tracer of the total mass of molecular gas. 
Additionally, the simultaneous observation of the \thCO\ and C$^{18}$O 
lines can provide useful chemical information on MCs. By combining 
multiwavelength data from molecular gas and other tracers of stars, 
we are well placed to investigate the star formation in the Milky Way.
So far, millimeter and submillimeter CO observations have been 
particularly important in studying molecular gas characteristics 
toward the Galactic CMZ.

Benefiting from the high-quality \twCO\ and \thCO\ data from the Milky
Way Imaging Scroll Painting (MWISP) project \citep[][]{2019ApJS..240....9S},
\cite{2024ApJ...971L...6S} have identified many CO structures with unique 
morphological and kinematic features distributed along the major axis of 
the Galactic bar \citep[see the Galactic stellar bar in, e.g.,][]
{2015MNRAS.450.4050W,2016ARA&A..54..529B,2017MNRAS.471.4323S,
2020RAA....20..159S,2023MNRAS.520.4779L,2024MNRAS.528.3576V}.
The bar channel traced by the CO dust lane connects the Galactic disk at 
the Galactocentric distance of $R_{\rm GC}\sim$~3.2--3.4~kpc and the CMZ 
at $R_{\rm GC}\sim0.3$~kpc. 
On the other hand, the length of the bar dust lanes is generally consistent with 
the radial scale length of the Galactic exponential disk \citep[i.e.,
$\rho(R)\propto e^{-R/R_{\rm sl}}$ and $R_{\rm sl}\approx3.2$~kpc,
see Figure~13 in][]{2021ApJ...910..131S}, likely indicating that the bar
dynamics have a significant impact on the properties of the molecular gas
in the inner Galaxy (e.g., distribution, morphology, kinematics, chemistry, 
and evolution). Our study provides solid evidence that the large-scale 
molecular gas inflow is driven by the Galactic bar toward the CMZ 
\citep[see additional observations and simulations; e.g.,][]
{1999A&A...345..787F,2008A&A...489..115R,2015MNRAS.449.2421S,2019MNRAS.484.1213S}.

The CO flows in the bar channel, driven by the nonaxisymmetric potential, 
become a secular and stable
gas reservoir that feeds star-formation in the CMZ and potentially supplies 
the supermassive black hole (SMBH) in the Galactic center (GC).
The gas inflow rate onto the CMZ is estimated to be about
$1.1\pm0.3\ \Msun$~yr$^{-1}$ in a transport period of $\sim$24~Myr
by adopting an average inflow velocity of $\sim140\km\ps$
for the large-scale gas lane \citep[][]{2024ApJ...971L...6S}. 
According to the MWISP CO data, the pattern speed of 
$\Omega_{\rm bar}\lsim32.5\pm2.5\ \km\ps\kpc^{-1}$ supports that 
the Milky Way is currently in a slow-bar scenario, which is consistent 
with recent studies \citep[e.g.,][]{2020MNRAS.497..933H,
2022MNRAS.512.2171C,2022ApJ...935...28W,2024MNRAS.533.3395Z}.

Despite much progress being made in multiwavelength observations
and numerical simulations, some open issues still need to be carefully
investigated so that we can gain a more accurate understanding of the
gas dynamics in the inner region of the Milky Way. How does the gas 
with significant angular momentum transfer from the Galactic disk 
several kiloparsecs away to the CMZ and even the GC? 
How are the molecular gas fuels that subsequently form stars affected 
by the Galactic large-scale dynamics? What is the relationship 
between the gas inflow rate and the efficiency of the star formation 
in the CMZ? And what role do massive stellar winds, photoionization, 
radiation pressure, and supernova feedback play in the gas accumulation 
of the CMZ? 

Our observations are instantaneous and freeze-frame. 
Can the observations accurately reveal the dynamical evolution of molecular 
gas flows toward the CMZ over time? In some specific studies, numerical 
simulations often simplify and discard many of the details of the
large-scale gas flows in which the CMZ is embedded. Can some of the striking 
and interesting features in the simulations be verified by observations?
In this paper, we investigate the dynamical nature of gas flows toward 
the CMZ based on the MWISP CO survey. Thanks to the large-scale MWISP 
data with high spatial resolution and sensitivity, we are able to 
capture many spatial and kinematic details of the molecular gas flow 
toward the CMZ that have never been revealed in previous studies.

In Section 2, we show that the tilted overshoot gas flows from the far 
dust lane converge in the near bar channel and eventually settle into 
the CMZ. In Section 3, a schematic diagram is constructed to illustrate 
both the overall dynamic behavior and detailed gas flow patterns toward the CMZ. 
We suggest that the dissipation processes resulting from collisions between 
large-scale gas flows are critical for removing the gas's angular momentum 
and facilitating its inward migration by considering the overshoot effect.
In Section 4, we propose that the nonaxisymmetric gravitational potential 
of the bar plays a significant role in regulating the gas 
dynamics within 3~kpc of the Galaxy during its secular evolution.
Finally, we give a summary in Section 5.

\section{Overshoot Gas from the Far Dust Lane}
In Figure~\ref{fig:f1}, the kinematic properties 
of the large-scale CO structures are well delineated toward the CMZ in the 
region of $l=[1\fdg2, 16\fdg2]$ and $v_{\rm {LSR}}=[-120\km\ps, +300\km\ps]$. 
Besides the CO inflow driven by the Galactic bar \citep[hereafter DL1; see 
the details in][]{2024ApJ...971L...6S}, 
two additional large-scale CO structures have been revealed flowing toward 
the CMZ along the near dust lane \citep[i.e., DL2 and DL3; also refer to][]
{2006A&A...447..533L,2008A&A...486..467L,2019MNRAS.488.4663S}. 
These gas flows are receding relative to the observer, so 
we identify them with red arrows in the $l$--$v$ diagrams.

Unexpectedly, a new CO structure with the most negative velocity is 
discovered below the Galactic plane according to the large-scale MWISP 
data (the long blue arrow in the lower panel of Figure~\ref{fig:f1}).
The new kinematic CO structure consists of many filamentary MCs
and exhibits a $v_{\rm {LSR}}$ gradient that increases with Galactic
longitude. Obviously, the CO feature is not related to the near 3~kpc
arm \citep[see the long blue line in the lower panel of Figure~\ref{fig:f1}; 
also refer to][]{2008Dame,2016ApJ...823...77R,Reid19,2025ApJ...982..185K}. 
On the contrary, the large-scale gas structure coexists with DL1 in a 
similar $l$--$b$ range. More precisely, the approaching flow is roughly 
parallel to or slightly below the near dust lane (i.e., see the colors 
for the approaching gas and the white contours for the receding 
DL1 gas in Figure~\ref{fig:f2}(c)). We define the large-scale approaching 
CO gas as the overshoot flows from the far dust lane \citep[e.g.,][]{2006A&A...447..533L,2022ApJ...939...58W}.

In Figure~\ref{fig:f2}(c), an intriguing filamentary cloud associated with 
the overshoot gas flow is found to be located in the G7 region
($l\sim7\fdg3, b\sim-1\fdg1$) with a large velocity gradient of 
$\gsim+200\km\ps\kpc^{-1}$ by adopting a distance of $\lsim$6~kpc
\citep[see the parameters of the geometry in][]{2024ApJ...971L...6S}. 
The value is at least 2 times larger than the mean velocity gradient 
of the near dust lane (i.e., $\sim -100\km\ps\kpc^{-1}$ for DL1), 
indicating rapid deceleration of overshoot gas in the G7 region. 
The feature stems from the fact that the overshoot gas is undergoing 
strong tidal processes when it moves in the Galactic bar potential. 
As molecular gas flows pass through the apocenter of elongated 
elliptical orbits, MCs experience extremely strong torques and shear forces, 
resulting in significant geometric deformation from intense tidal effects 
\citep[e.g.,][]{2019MNRAS.484.5734K}. 
Moreover, a considerable amount of overshoot gas is being pulled 
back by the Galactic bar potential and reenters the bar channel toward 
the CMZ, which is well delineated by the short blue arrows (OG1 and OG2) 
and the red arrows (DL2 and DL3) in Figure~\ref{fig:f1}. 

The overshoot flows are changing their trajectories in response to 
the Galactic bar potential, i.e., (1) from OG1 below the Galactic plane to DL2 
above the plane (the upper panel of Figure~\ref{fig:f1}) and (2) from OG2 
at $b\sim-1\fdg9$ to DL3 at $b\sim0^{\circ}$ (the lower panel of 
Figure~\ref{fig:f1}). Indeed, we distinctly illustrate that the 
large-scale multilayer shock structures with varying velocities 
are located in coherent but slightly different positions, indicating that
the motion state of a series of overshoot flows with similar trajectories
has changed dramatically in the vicinity of this region (see the shock 
structures indicated in blue, red, and green colors close to the G5 
region at $l\sim5\fdg4$ and $b\sim-0\fdg4$; Figure~\ref{fig:f2}(a)). 
The data show that the overshoot gas is reaccelerating from 
$v_{\rm {LSR}}\sim-40\km\ps$ to $v_{\rm {LSR}}\sim+180\km\ps$ and 
even more at about half of the Galactic bar (i.e., at the middle
of the near dust lane DL1 of $l\sim5\fdg7-4\fdg9$ or 
$R_{\rm GC}\sim$~1.7--1.5~kpc) when it passes through the bar channel. 
The observations appear to be broadly consistent with the scenario
of significant gas accumulation around the inner Lindblad resonance
\citep[ILR; e.g., $R_{\rm ILR}\sim$~1.6~kpc in][]{2024MNRAS.528.5742S}.

Collisions between gas flows are also revealed by observations. 
Combined with complementary distributions between the high-velocity gas 
(i.e., white contours for the DL2 inflows in Figure~\ref{fig:f2}(b)) and the 
low-velocity overshoot gas \citep[colors for the OG3 flows; also see Figures~6 
and 7 in][]{2020MNRAS.498.5936E}, for example, the prominent CO protrusions 
evidently demonstrate that the two gas flows in opposite directions are 
colliding violently at the G3 region of ($l\sim3\fdg2$, $b\sim0\fdg3$).
Interestingly, the receding gas of DL3 passing below G3 (see red color near 
$l\sim3\fdg4$--$2\fdg9$, $b\lsim0\fdg2$--$0\fdg0$ in Figure~\ref{fig:f2}(b)) 
also appears to collide with the surrounding overshoot gas, causing G3 to exhibit 
a deformed structure at ($l\sim3\fdg05$, $b\sim0\fdg15$).
The observations are not only clear but also 
align well with the cloud--cloud collision scenarios discussed in 
the literature \citep[e.g., see][]{2009ApJ...696L.115F,2013ApJ...774L..31I,
2017ApJ...835L..14G,2021PASJ...73S..75E,2021PASJ...73S...1F,2022MNRAS.514..578W,
2023ApJ...959...93G}. According to the large-scale bow shock perpendicular to 
the dust lane in Figure~\ref{fig:f2}(a), the G5 region at $R_{\rm GC}\sim$1.6~kpc 
is another excellent example of the collision between gas flowing along 
the bar channel (i.e., white contours for high-velocity receding gas in DL1) 
and the approaching overshoot flows passing through the bar channel 
\citep[i.e., the blue and green colors for the low- and middle-velocity 
gas in OG2+DL3; also see][]{2021MNRAS.502.5896A}.
We believe that the approaching gas flows from overshoot are bombarding 
the large-scale receding gas streams in the vicinity of the near dust lane.

It is noteworthy that the large-scale shock structure near G5 (indicated by the yellow 
dotted line in Figure~\ref{fig:f2}(a)) spans approximately 2\fdg7 on the sky plane, 
corresponding to a projected physical size of about 320~pc at the distance of 6.8~kpc
\citep[see the geometry of the bar lanes in][]{2024ApJ...971L...6S}.
Here, the size of $\sim$320~pc may provide a reliable measurement of the
cross-sectional diameter of the bar channel, which is comparable to the size of 
the CMZ \citep[e.g., see][]{2023ASPC..534...83H}. Indeed, the shock features 
(i.e., velocity jump) on the bar region can be well reproduced in simulations 
\citep[e.g.,][]{1992MNRAS.259..345A,2012ApJ...751..124K,2024ApJ...963...22F}.
And our observations can provide additional constraints for hydrodynamic 
simulations of the Milky Way case.
Potential signs of star formation are also observed near the apex boundary 
of the large-scale shock structure \citep[see GMC M4.7-0.8 in][]{2025arXiv250314174B}.
This result may support the hypothesis that as the gas flow approaches the orbital 
apocenter (this work), the decrease in orbital velocity induces gas compression 
(density enhancement), thereby triggering localized star formation 
\citep[][]{2025PASA...42...14C}.
We propose that the strength of the inflowing gas, as well as gravitational 
potential (e.g., bar strength, axis ratio, and central mass distribution) and 
local interstellar medium properties (e.g., turbulence level, magnetic field 
strength, and star formation activity), influence the size and other properties 
of the Galactic CMZ. 

At least two small breaks along DL1 can be identified in the 
$l$--$v$ diagram (see the tips of OG1 and OG2 in the lower panel of 
Figure~\ref{fig:f1}). The features can naturally be explained by 
the convergence of the overshoot gas from the far dust lane to 
the DL1 fresh gas from the $R_{\rm GC}\sim$3--4~kpc region, 
leading discernible changes in the velocity gradient and gas 
distribution of the whole gas inflows along the near dust lane. 
The scenario is also consistent with the observational fact that 
lower-velocity overshooting gas is mainly concentrated upstream, 
while the-higher-velocity DL1 gas becomes enhanced downstream 
(i.e., the slight displacement between the overshoot 
gas represented by the colors and the DL1 gas by white contours 
in Figure~\ref{fig:f2}(c)). Moreover, the gradual velocity-shift
features are indeed discerned along and/or near OG1 and OG2 in
the $l$--$v$ diagrams (Figure~\ref{fig:f1}). For the first time,
our observations provide detailed evidence that large amounts of
overshoot gas from the far dust lane are captured by the Galactic 
bar potential at these different $R_{\rm GC}$ and then 
inflow toward the CMZ along or parallel to the bar channel 
(e.g., DL2 and DL3).

According to the Gaussian decomposition and clustering algorithm
\citep[e.g., see][]{2019MNRAS.485.2457H,2019A&A...628A..78R,2024AJ....167..220Z},
we identified the MCs toward the CMZ by using the smoothing parameters of
$\alpha$1=1.8 and $\alpha$2=5.4 for the MWISP \twCO\ data with the 3D grid
of $0.5'\times0.5'\times0.5\km\ps$. It is straightforward to pick out
the DL1 samples from the near dust lane and the overshoot samples from
the far dust lane based on the $l$-$v$ diagrams (Figure~\ref{fig:f1}). 
Accordingly, the intermediate samples are selected from MCs with 
$v_{\rm {LSR}}$ between the DL1 and the overshoot samples, i.e., 
MCs in DL2, DL3, G3, and G5 and near the head part of OG1, OG2, and OG3, 
excluding those from the blended regions.
Table~1 displays the statistical properties of three types of MCs
with angular sizes larger than 50~arcmin$^2$.

By employing the method consistent with our earlier work \citep[][]
{2024ApJ...971L...6S}, we consistently derive the CO-to-H$_2$ conversion
factor of $X_{\rm CO}\sim0.7-1.2\E{20}$~cm$^{-2}$(K~km~s$^{-1})^{-1}$.
The value is about half of the commonly used value \citep[e.g.,][]
{2001ApJ...547..792D,2004A&A...422L..47S,2013ARA&A..51..207B,2024MNRAS.527.9290K}
and close to $\sim(0.5-1.0)\E{20}$(K~km~s$^{-1})^{-1}$ for the Galactic
bulge region \citep[e.g.,][]{2014A&A...566A.120S}.
Due to shocks driven by the bar and frequent cloud--cloud collisions,
there may be multiple gas components with different temperatures and
densities near the bar channel \citep[e.g.,][]{2024ApJ...977...37N},
resulting in systematic changes in the molecular gas state.
Generally, in the strong shock and collision regions of G3 and G5,
MCs with wide line broadenings often
have a relatively small ratio between \thCO\ and \twCO\ intensity,
which is about half or less of the value of the normal MCs in the Galactic plane 
\citep[e.g., $I(^{13}$CO)/$I(^{12}$CO)$\sim$0.2; see][]
{2019PASJ...71S..19T,2023AJ....166..121W}. The same is true of statistics 
for other gas flows associated with the Galactic bar (i.e., the large-scale 
MC samples with less contamination in the near dust DL1 and overshoot flows; 
see Table~1). Based on observations, the statistical result of the 
smaller $I(^{13}$CO)/$I(^{12}$CO) likely represents the 
lower $X_{\rm CO}$ in the complex shock environments of the inner Galaxy.

The Milky Way represents an excellent case study for understanding gas 
cycling in galaxy evolution. Detailed observations have provided a wealth of 
data and information about the Milky Way, such as the 3D spatial distribution 
of gas flows, the precise geometric structure of the bar, and other relevant
key physical parameters (e.g., the gas inflow rate, the outflow rate, the
star formation rate in the CMZ, etc.).
These detailed observational insights are crucial for constructing and 
constraining theoretical models, as they significantly advance our 
understanding of gas dynamics in galaxies and their secular evolution.
Intriguingly, external galaxies also exhibit similar overshooting features 
\citep[NGC 1097 and NGC 1365; see, e.g.,][]
{2023MNRAS.523.2918S,2023ApJ...944L..14W,2023ApJ...944L..15S}. 
Exploring the gas cycling processes and associated physical parameters 
(e.g., geometry of the bar and overshooting gas, distribution of gravitational 
potential, gas inflow rate, etc.) in external galaxies and comparing
them with those in the Milky Way can elucidate the fundamental role of gas 
cycling in the broader context of galaxy evolution.

In summary, the molecular gas is a good indicator of the 
gravitational potential of the Galactic bar because the gas flows 
can be easily and quickly affected by the change in the gravitational 
torque of the rotating bar. The dynamical processes dominated by the 
gravitational potential of the bar cause the overshoot gas to 
decelerate, accumulate, and collide violently at certain locations, 
systematically changing the gas motion and its overall distribution 
toward the CMZ. We propose that gas dynamics play a crucial role 
in studying several key issues, such as the redistribution of inflowing 
material and the regulation of overall gas inflow efficiency (Section 3),
as well as the subsequent star formation efficiency in the CMZ (Section 4).

\section{Overall Dynamics of the Gas Flows toward the CMZ}
Using new data from the MWISP and supplements from other CO surveys 
\citep[e.g.,][]{2001ApJ...547..792D,2019PASJ...71S..19T,
2020MNRAS.498.5936E,2021MNRAS.500.3064S}, 
we construct a sketch to illustrate the overall dynamical nature and 
details of molecular gas flows toward the CMZ.
The upper portion of Figure~\ref{fig:f3} displays an edge-on view of 
the spatial distribution and motion characteristics of the overall 
gas flows from the perspective of the observer,
while the lower portion shows gas flows dominated by the Galactic bar's 
gravitational potential in a face-on view. For the edge-on view, note that
the gas inflows extend to $l\sim+16^{\circ}$ in the positive longitude
\citep[i.e., the near dust lane DL1; see][]{2024ApJ...971L...6S} 
and $l\sim-8^{\circ}$ in the negative longitude for the far dust 
lane by assuming that it has a similar length and inclination angle of 
$\phi_{\rm bar}=23^{\circ}\pm3^{\circ}$ 
(between the bar major axis and the Sun-GC) as DL1. 
The realistic scenario may be much more complex than 
what we depict here, but this sketch helps us get a comprehensive picture 
of the entire gas dynamics process toward the CMZ.

Many of the physical processes in the CMZ, including gas dynamics,
have been studied in simulations, in which gas is compressed
and shocked as a result of the shift from x1 orbit family to x2
orbit family \citep[e.g.,][]{1992MNRAS.259..345A,1999MNRAS.304..512E,
2002MNRAS.329..502M,2004ApJ...600..595R,2005ApJ...632..217S,
2008A&A...489..115R,2014A&A...565A..97C,2015MNRAS.449.2421S}.
As seen in Figure~\ref{fig:f3}, our observations align well with the model 
in which the Galactic bar potential 
(1) channels a substantial amount of molecular gas from the inner Galactic 
disk at $R_{\rm GC}\sim$3.2--3.4~kpc toward the CMZ (i.e., DL1 and the 
associated kinematic features of CO gas), 
(2) generates noncircular motions and large-scale shock structures along 
the near dust lane \citep[e.g., a series of bow shocks revealed by CO gas;]
[]{2024ApJ...971L...6S}, 
(3) governs and regulates the formation and evolution of the CMZ through 
the overshoot mechanism (i.e., gas flow collision/accumulation near G3/G5/G7 
and multiple inflows DL2/DL3 with different trajectories toward the CMZ 
in Section 2; also see details in the following discussion), and 
(4) has important impacts on the gas distribution, properties, and 
subsequent star formation within the inner Galaxy (e.g., the 3~kpc arm, 
the thickness of the gas disk and the star formation rate therein, etc.). 
As another good example of gas influenced by the nonaxisymmetric bar potential, 
we will further discuss the tilted bar lanes traced by CO flows in Section 4.

The Galactic bar systematically transports a massive inflow 
toward the CMZ. However, this is a necessary but not a sufficient
condition for the accumulation of gas in the CMZ.
Another key question is how to effectively remove the large amount
of angular momentum of the gas located a few kiloparsecs away, so that the
gas can settle continuously into the region within
the central few hundred parsecs of the Milky Way. 
Next, we will focus on how the overshoot effect plays a critical role 
in regulating the redistribution of gas and its angular momentum, 
as robustly evidenced by several key kinematic characteristics and 
dynamical effects observed in the new CO flows toward the CMZ.

In Section 2, we undoubtedly confirmed that G3 originates in violent 
collisions between the inflowing gas of DL2/DL3 and the overshoot gas 
from the far dust lane (the G3 case for x1 orbits intersects with x2 orbits; 
Figure~\ref{fig:f2}(b)). In this way, G3 is located near the outskirts of 
the CMZ (i.e., at $R_{\rm GC}\sim$~0.5~kpc). Meanwhile, gas flow collisions 
also occur in regions where the overshoot gas changes its trajectory and 
passes through the bar channel due to varying torques. When the overshoot 
gas enters the bar channel, collisions between gas flows are inevitable at 
the intersections of different eccentric orbits (the G5 and G7 cases for gas 
flow interactions in bar channel; Figures~\ref{fig:f2}(a) and (c)).
Therefore, collisions between gas flows are very common in extremely crowding orbits, 
especially in the region where the x1 orbits of gas intersect with the x2 orbits 
in nonaxisymmetric gravitational potential (Figure~\ref{fig:f3}). 

The starting points of DL2 and DL3 are located exactly around 
the regions of G7 and G5, respectively. And the gas mass in the intermediate 
samples is 1 order of magnitude higher than that in the overshoot gas, 
confirming significant accumulation of the overshoot gas within the bar channel 
at these certain locations (e.g., near the G7, G5, and G3 regions).  
The consistency may indicate that, due to the dynamics of the rotating bar,
a significant accumulation of overshoot gas occurs in some special regions 
\citep[i.e., near G5 at $R_{\rm GC}\sim\frac{1}{2}R_{\rm bar}$ 
and G7 at $R_{\rm GC}\sim\frac{2}{3}R_{\rm bar}$ by adopting the bar length 
of $\sim$3.2--3.4~kpc; see][]{2024ApJ...971L...6S}.
In addition, considerable CO emission in DL1 is also enhanced significantly 
near these regions \citep[$l=4\fdg8-5\fdg6$ and $l=6\fdg9-8\fdg3$
in Figure~\ref{fig:f2}(c); also see Figure~1 in][]{2024ApJ...971L...6S}.
These regions are located near the apocenter of the overshoot flows 
(Figure~\ref{fig:f3}). The interesting feature needs to be further explored.

Importantly, the measured line-of-sight velocities of the DL2 and DL3
inflows toward the CMZ are $\sim$1.5--8 times smaller than those of
the DL1 at a similar $R_{\rm GC}$ \citep[see Figure~\ref{fig:f1};
also refer to the geometry proposed in][]{2024ApJ...971L...6S}.
The observations clearly indicate that the tangential velocity of gas rotating
clockwise around the GC has significantly decreased, and the orbital angular
momentum of overshoot gas flows has been reduced by approximately 0.2--0.7~dex
after being captured and reaccelerated by the bar potential.
These features are consistent with the fact that there is substantial
exchange of mass and angular momentum as overshoot gas changes from approaching
flows to receding flows at some locations (e.g., the G7 region at $\sim$2.2~kpc,
the G5 region at $\sim$1.6~kpc, and the G3 region at $\sim$0.5~kpc;
see Figures~\ref{fig:f2} and \ref{fig:f3}). Notably, our observations provide
the first clear evidence that the overshoot gas, after being captured by the
bar potential during its passage through the bar channel, continues to flow
toward the CMZ with significantly reduced angular momentum.

The gas mass in DL1 and that associated with the overshoot effect are
$\sim0.9\times10^{7}\ \Msun$ and $\gsim1.8\times10^{7}\ \Msun$ (Table~1), respectively.
The result indicates that most of the accretion gas does not enter the CMZ
directly but rather overshoots into the bar channel on the other side.
Based on our observations, the current inflow efficiency can be easily
obtained to be $\sim$1/3 on average, which agrees well with the simulation
\citep[e.g.,][]{2021ApJ...922...79H}.
Considering the importance of the overshoot effect \citep[e.g.,][]
{2002MNRAS.329..502M,2019MNRAS.484.1213S,2021ApJ...922...79H}, we estimate
an average inflow rate as $\gsim1/3\times2\times(0.9\E7\ \Msun/24\ {\rm Myr}+1.8\E7\ \Msun/15\ {\rm Myr})\gsim1.1\ \Msun\yr^{-1}$.
Here, 1/3 is the observed inflow efficiency, while 2 corresponds to
both the near and far bars. 
And 15~Myr comes from the gas kinematics of the DL2 and DL3 
inflows by adopting an average inflow velocity of $\sim90-100\km\ps$ 
(see DL2 and DL3 in Figure~\ref{fig:f1}).
Note that the instantaneous accretion rate
can be several times higher than the average value.
For example, it can easily cause the accretion rate to increase by nearly
five times as several MCs (e.g., with a mass of $\sim10^5\ \Msun$ and
a size of $\sim10$~pc) with low angular momentum (and thus higher inflow
efficiency than 1/3) settle into the CMZ within $\lsim$0.1~Myr.
The gas accumulated in the G3/G5/G7 regions serves as a potential
reservoir to regulate the instantaneous gas inflow rate.

Again, not all bar-driven gas flows directly enter the CMZ;
instead, they take approximately three orbital periods to settle into the CMZ 
(i.e., the observed small inflow efficiency of $\sim$1/3). Through this
process, the gas is subjected to intense gravitational torques from the bar, 
causing it to be stretched, distorted, shredded, torn apart,
and fragmented (Figure~\ref{fig:f2}(c)).
Apparently, the overshoot effect greatly lengthens the action time
of various dynamical processes before the accretion gas enters the CMZ.
Throughout the entire dynamical process, the angular momentum of 
the gas changes and redistributes due to the combined effects such as torque 
interactions from the bar, inelastic cloud--cloud collisions, gas viscosity, 
and various instabilities. The shock dissipation generated during 
the sufficient dynamics will redistribute angular momentum outward. 
This enables the distorted gas flows to efficiently transport material 
from kiloparsec scales to regions within a few hundred parsecs of the GC
while also supporting the scenario of the periodic migration of gas toward the CMZ.

Briefly, the confirmation of the overshoot gas, together with 
frequent collisions caused by intersecting flows, provides detailed 
observational evidence to explain the dynamics of gas flows toward the CMZ.
The overshoot gas with lower angular momentum gradually converges 
into the bar channel, then flows toward the CMZ again 
(see DL2 and DL3 in Figure~\ref{fig:f3}).
Because of the overshoot effect and frequent collisions, the accretion gas 
is able to efficiently transfer its excess angular momentum outward through 
various dynamical processes and then cause the gas to gradually flow 
toward the CMZ. Here, the gravitational torque of the robust rotating 
Galactic bar makes a large contribution in the process of redistributing 
the gas and removing angular momentum from the molecular gas that 
eventually settles into the CMZ. Due to the interaction of various torques 
with viscous gas and the frequent collisions between gas flows, 
at the same time, the strong DISSIPATION process is also essential 
for the effective outward transfer of gas angular momentum, 
which in turn remarkably regulates the gas flow dynamics and gas 
accumulation in the CMZ over the secular evolution of the Galaxy.

\section{Large-scale Tilted Gas Flows and the Twisted $\infty$ Structure in the CMZ}
The tilted dust lane has been discussed in many works
\citep[e.g.,][]{2006A&A...447..533L,2008A&A...486..467L,2006A&A...455..963R,
2008A&A...477L..21M,2019MNRAS.484.1213S,2020MNRAS.499.4455T,2021MNRAS.502.5896A}.
We also confirm that the near dust lane DL1 with a length of 
$\sim3.2$--3.4~kpc exhibits distorted features during the large-scale gas
accretion driven by the Galactic bar. Additionally, DL1 is distinctly located 
below the Galactic plane, which is exactly the opposite of the gas flow of 
the far dust lane above the plane (see Figure~\ref{fig:f3}). 

Our previous studies showed that the slightly tilted midplane to the 
IAU-defined plane \citep[i.e., $b=0^{\circ}$, see Figure 7 in][]
{2016ApJ...828...59S} is caused by the Sun's height above the physical 
midplane of the Milky Way \citep[i.e., $z_{\rm sun}\sim$15--17 pc;
see][]{2016ApJ...828...59S,2019ApJS..240....9S,2021ApJ...910..131S}.
However, the tilted large-scale gas flows toward the CMZ cannot be 
explained by this \citep[e.g., see discussions in][]{2016ARA&A..54..529B} 
because the maximum deviation of the CO gas structure reaches 
$\lsim -$170~pc (or $b\lsim -1\fdg5$ at $l\sim +5\fdg5$) from 
the Galactic plane of $b=0^{\circ}$.
The deviation value is about 2 times larger than the thickness of the thin 
CO disk \citep[see Table 3 in][]{2021ApJ...910..131S}. Considering the 
smaller thickness of the molecular gas disk in the inner 3--4~kpc region, 
this deviation from the $b=0^{\circ}$ plane is even more severe.
We attribute the large-scale gas flows to a true tilted feature relative 
to the Galactic midplane. 
Based on our CO data and model \citep[][]{2024ApJ...971L...6S}, 
the tilted angle of the bar lanes relative to the Galactic midplane in 
$l=[+6^{\circ},-4^{\circ}]$ is estimated to be $\theta_{\rm{bar\ lanes}}\gsim5^{\circ}$, 
which roughly agrees with the simulations \citep[e.g., $\sim0^{\circ}-5^{\circ}$ in][]
{2020MNRAS.499.4455T}. 
Indeed, the gas structure revealed by the \HI\ emission exhibits 
a more extreme tilted characteristic \citep[i.e., $b\lsim -3\fdg5$ or 
$z_{\rm gas}\lsim -$350~pc for the atomic gas with the 
most negative velocity; see][]{1983A&AS...52...63B}.

The vertical distribution of gas (e.g., the tilted and warped gas flows) is 
modulated by the strong gravitational torque generated by the rotating bar and 
various dissipative forces. In our observation, the gravitational potential of 
the bar can effectively trap the gas flows into the dust lane (Figure~\ref{fig:f3}). 
The gas on different orbits is perturbed by 
the nonaxisymmetric gravitational potential, 
leading it to undergo noncircular motion around the GC and oscillations 
relative to the bar. Due to the varying torque exerted by 
the bar's potential on the gas, gas flowing along different trajectories 
and inclinations will move with varying precession rates until it eventually 
settles into the bar's potential, aligns with the bar's major axis, 
and then rotates around the GC with the bar. 
These torques also induce misalignments across different parts of the 
bar lanes. 
When gas flows on different orbits are subjected to varying torques, 
they develop distorted or warped structures as observed in CO data. 
In the process, the angular momentum of the overshoot flows is also 
transferred to the bar through the dynamic friction.
Alternatively, the distortion of the bar lanes caused by resonant trapping 
of x1 orbits \citep[e.g.,][]{2019A&A...629A..52L} may also influence the
distribution and motion of the gas.
The ongoing interaction between the gas flows and the bar itself accumulates 
over time, continuously affecting the bar's structure and motion as it
gains mass and angular momentum by exerting torque on the gas.

The above processes are accompanied by complex dynamical interactions
that alter the angular momentum and motion states of the gas flows.
We confirmed that collisions between gas flows are frequent 
during the dynamical processes in the G3, G5, and G7 regions.
On the one hand, the nonaxisymmetric bar decouples the gas inflows from 
their initial dynamical state, resulting in rapid assimilation of the accretion
gas into the bar potential (e.g., DL2 and DL3).
On the other hand, the frequent collisions between gas flows (G3 and G5),
combined with the tangential shear of the tidal forces exerted by the 
nonaxisymmetric bar, result in significant dissipation in the twisted gas flows.
Indeed, the typical timescale of these dissipative processes,
which are driven by shocks and strong turbulence, may be comparable
to the short dynamical timescales associated with gas flow collisions.
During the dynamical process of overshoot gas settling into the
Galactic bar potential, the dissipation from collisions between 
misaligned gas flows, together with gas viscosity and the effects 
of various torques, serves as an efficient mechanism for the outward
transfer of angular momentum of gas, thus facilitating their 
gradual migration toward the CMZ.

The scenario is strongly supported by our observations that the 
overshoot gas is not fully relaxed into the Galactic bar potential. 
That is, the overshoot gas from the far dust lane exhibits a more 
tilted structure (i.e., $\sim$1.5 times larger; see $b\lsim -1\fdg8$ 
from $l\sim +3^{\circ}$ to $+11^{\circ}$ in Figure~\ref{fig:f2}(c)) 
than the near dust lane DL1 below the Galactic plane, while a portion 
of the gas passing through the bar channel even extends above 
the Galactic plane (i.e., DL2 is above and almost parallel to the 
Galactic plane). The distorted or warped gas structures in observations 
may represent the oscillatory characteristics of gas flows perpendicular 
to the Galactic bar as it moves and relaxes in the bar potential.

Meanwhile, the motion state of the self-consistent bar evolves 
over time through the exchange of angular momentum between the bar 
itself and the accreted material. As a result, the bar gains angular 
momentum from the captured gas and loses it to the halo and the outer 
stellar disk \citep[e.g.,][]{2021A&A...656A.156Q,2022A&A...659A..80W,
2023MNRAS.524.3596D}, leading to a gradual change in the state of the 
bar \citep[e.g., the morphology, structure, and kinematics; see][]
{2016ApJ...824...13L,2021MNRAS.500.4710C,2022MNRAS.512.2171C,
2023MNRAS.520.4779L,2024MNRAS.528.3576V}. We suggest that the 
nonaxisymmetric bar 
is essential for regulating the redistribution of accreted gas 
and its angular momentum transfer. This self-regulating process may stably 
maintain large-scale gas accretion and gas recirculation over a 
considerable timescales in the presence of the bar \citep[e.g., 
$\sim$4--8~Gyr, see][]{2019MNRAS.490.4740B,2024A&A...681L...8N}.

The accumulated gas in the CMZ inherits the information of the gas 
supplied by the large-scale inflows, which includes the fresh gas 
from the disk at $R_{\rm GC}\sim$~3--4~kpc (with higher angular 
momentum but lower accretion rate) and the recycled gas from the
overshoot effect (with lower angular momentum but a higher accretion rate).
The moving direction of the large-scale inflows (DL1) is well consistent 
with the trend of the $\infty$ structure at the far side of
the CMZ (see the red solid arrow in the CMZ of Figure~\ref{fig:f3}).
That is, gas along the DL1 moves from the negative latitude at 
$l\sim+0\fdg8$ toward the positive latitude at $l\sim-0\fdg4$, 
and then back down toward the negative latitude \citep[e.g., see][]
{2011ApJ...735L..33M,2024arXiv241017321L,2024A&A...689A.121V,2024arXiv241017320W,
2025ApJ...982L..22K}. 

Simultaneously, the inflowing gas along the far dust lane moves from the positive 
latitude to the negative latitude at $l\sim-0\fdg6$, then toward the 
positive latitude at $l\sim+0\fdg7$, and finally back down toward the 
negative latitude again (i.e., the near side of the CMZ; see the blue solid 
arrow in the CMZ of Figure~\ref{fig:f3}). Similar to the previous discussion, 
when the gas flows pass near the x2 orbit, not all gas directly enters the CMZ; 
instead, some gas experiences the overshoot effect because of its higher angular 
momentum. Our data have clearly demonstrated that the overshoot gas flew past 
the near side of the CMZ and then collided violently with the inflowing gas 
from the DL2 and DL3 \citep[i.e., see the G3 region in Figure~\ref{fig:f2}(b); 
also see clouds studied by][]{2022A&A...668A.183B}.

We thus propose that the twisted $\infty$ feature of the CMZ
\citep[e.g.,][]{2011ApJ...735L..33M,2015MNRAS.447.1059K,2017MNRAS.469.2251R} 
arises from the combined effects of the large-scale tilted gas flows 
in the dust lanes and the locally distorted gas structure dominated by 
the strong gravitational potential within a few hundred parsecs of GC 
\citep[i.e., the nuclear stellar disk; see][]{2022MNRAS.512.1857S}.
The orbit periods of the twisted $\infty$ structure of the CMZ are about
2--4~Myr \citep[][]{2011ApJ...735L..33M,2015MNRAS.447.1059K},
suggesting a rapid precession of the gas under the influence of
external nonaxisymmetric gravitational potential and the
large-scale tilted gas inflows \citep[e.g.,][]{2020MNRAS.499.4455T}.
The timescale of several million years mentioned above is consistent with 
the dynamical time (2--3~Myr) of gas inflow from the G3 region into the CMZ,
indicating that external gas input significantly shapes the geometric
configuration and dynamic properties of the CMZ (e.g., local gas accumulation, 
twisted structure, and rapid precession).
Recent observations have revealed that the molecular circumnuclear 
disks with different kinematics are misaligned with the large-scale 
gas structure \citep[][]{2019A&A...623A..79C,2019MNRAS.489.3739R,
2020MNRAS.499.5719R,2023A&A...679A.115P}, indicating that highly warped 
gas structures are likely prevalent in the central region of galaxies.

The dynamical processes in the inner Galaxy cause gas flows to exchange 
mass and angular momentum with their surroundings, enabling the CMZ to gain 
fresh gas from the Galactic disk and the recycled gas from the overshoot effect. 
Given the long-term accretion process and the remarkable transient accretion 
effects caused by overshooting, the nonuniform accumulation of gas (e.g., the 
instantaneous accretion rate of about 5--10~$\Msun\yr^{-1}$) 
can significantly feed and regulate star formation in the CMZ.
Due to the highly distorted and inhomogeneous nature of gas inflows, 
the accretion process onto the CMZ, and even onto the SMBH, is likely highly 
nonuniform. This could result in the $\infty$ ring of the CMZ and jets of
the SMBH exhibiting strong precession and intermittent variability.
Furthermore, feedback from star-formation in the CMZ and the SMBH
\citep[e.g.,][]{2015MNRAS.446.2468E,2015MNRAS.447.1059K,2017MNRAS.466.1213K,
2018MNRAS.477.2716K,2019MNRAS.490.4401A,2020ApJ...901...51A,2020MNRAS.497.5024S,
2020MNRAS.499.4455T,2024A&A...685L...7M,2024ApJ...972L...3T} 
contributes to reducing the angular momentum of gas in the CMZ, 
enabling a fraction of lower angular momentum gas to 
migrate further toward the region closer to the GC and thereby 
increasing the potential activities of the SMBH. 

In simple terms, the gas and star-forming environment in the CMZ 
can be regulated by both the external gas input through the
bar channel (this work) and the various feedback therein
\citep[e.g.,][]{2019Natur.573..235H,2021A&A...646A..66P,
2024ApJ...966L..32M,2024A&A...681L..21N,2024A&A...691A..70N,
2024A&ARv..32....1S}, thus achieving a delicate 
balance between gas fuel supply and star formation in the extreme 
environments. Under the control of nonaxisymmetric gravitational 
potential of the Galactic bar, the continuous accumulations of gas with VARYING 
inflow rates, along with frequent interactions between gas flows 
and star formation, can exist simultaneously within a few hundred 
parsecs of the GC over a long timescale. This is what distinguishes 
the CMZ from other star-forming regions on the Galactic disk.

\section{Summary}
Based on CO morphological and kinematic details revealed by the MWISP data,
we investigate the nature of the approaching and receding gas flows toward
the CMZ. Our main results are as follows. \\

\begin{figure}
\vspace{-22ex}
\gridline{
  \hspace{-8.5ex} \fig{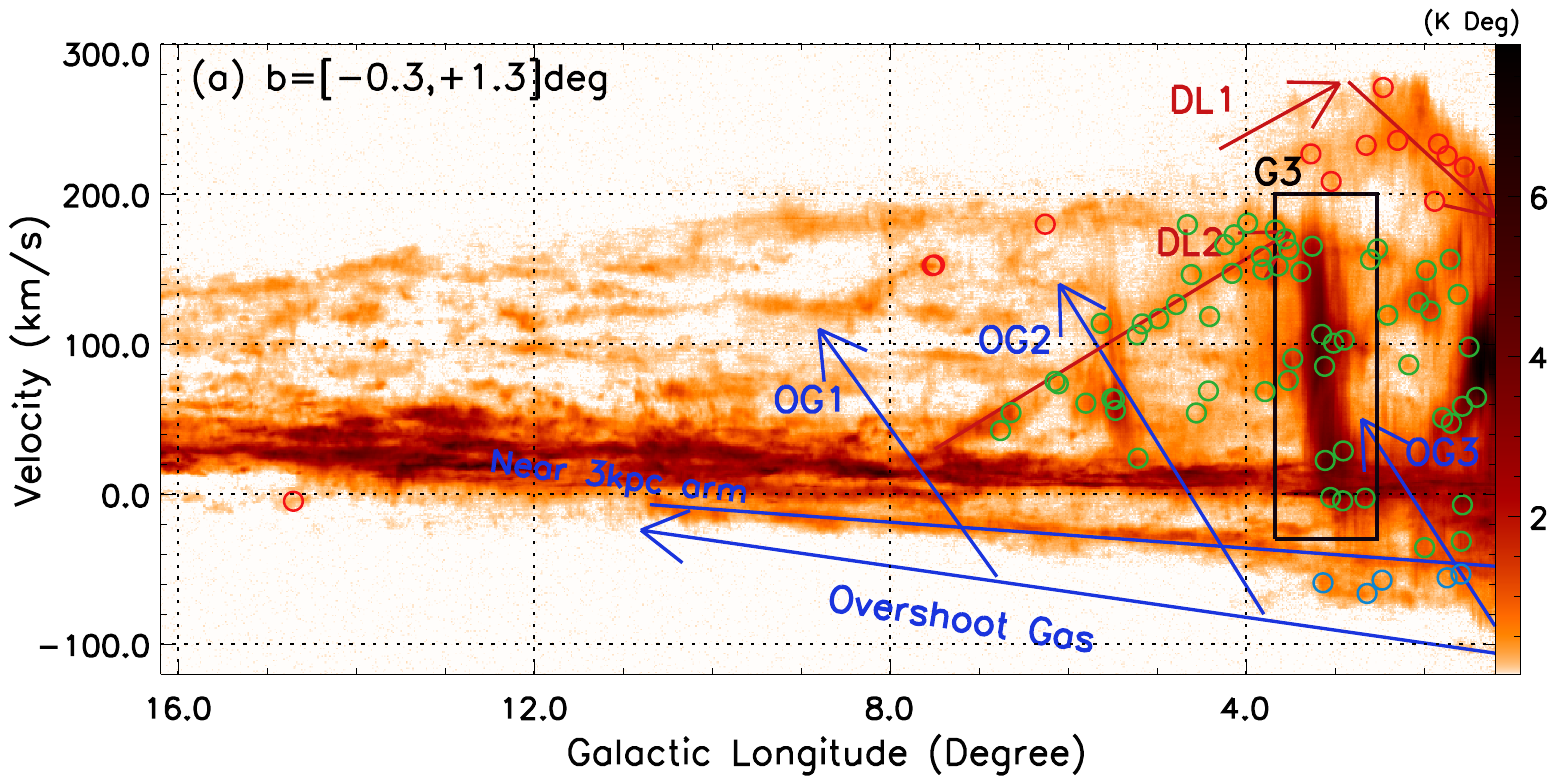}{1.2\textwidth}{} 
         }
\vspace{-42.5ex}
\gridline{
  \hspace{-8.5ex} \fig{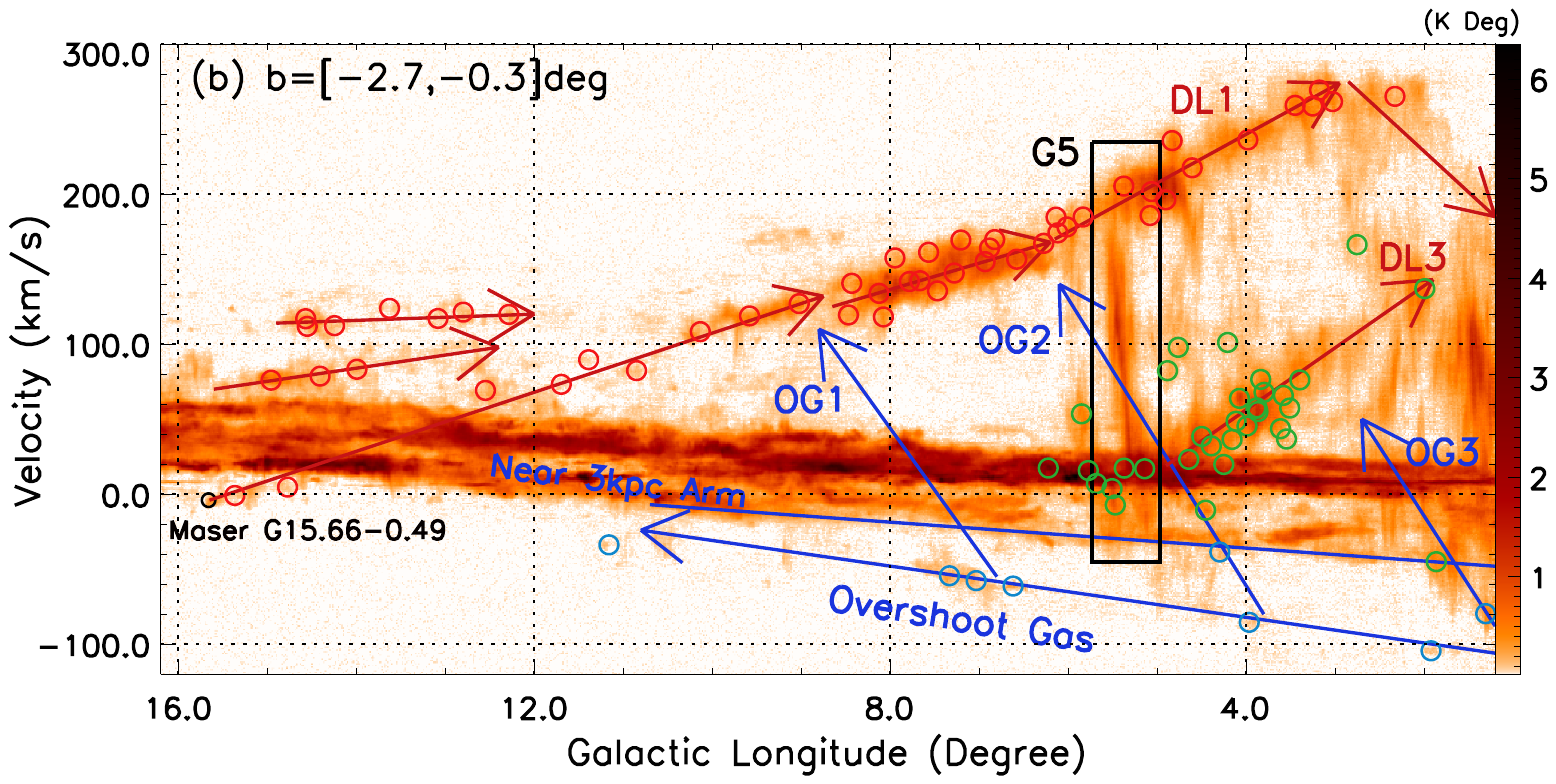}{1.2\textwidth}{} 
         }
\vspace{-22.5ex}
\caption{
Longitude--velocity diagram of \twCO\ emission from the MWISP survey 
\citep[][]{2019ApJS..240....9S} on a $0.5'$ grid along $l=16\fdg2$ -- $1\fdg2$.
The upper panel shows the $l$--$v$ features averaged over  
$-0\fdg3\leq b \leq  +1\fdg3$, while the lower panel displays the
main structure of the near dust lane over $-2\fdg7 \leq b \leq  -0\fdg3$.
The markers represent the different parts of the near dust lane 
(red arrows for main near dust lane DL1, and two other components of DL2 and DL3), 
the overshoot gas from the far dust lane (blue arrows), and the near 
3~kpc arm \citep[blue line; see text and also refer to Figure 3 in][]
{2024ApJ...971L...6S}. The rectangles show the extended velocity features of G3 
\citep[or Bania 2 in][]{1986ApJ...306L..17S} and G5 \citep[][]{2023ApJ...959...93G}.
The red, green, and blue circles represent the three categories of MC samples
with an angular size $>50 \ {\rm {arcmin}}^2$ as listed in Table 1.
The maser G015.66-00.49 \citep[at a parallax of 0.22$\pm0.029$~mas
and $v_{\rm {LSR}}=-4\km\ps$;][]{Reid19} is also labeled in the figure.
\label{fig:f1}}
\end{figure}

1. The MWISP CO data show that a large-scale gas structure, characterized by
highly negative velocities, is composed of many filamentary MCs and is moving
toward the observer (Figure~\ref{fig:f1}). Based on the observations, 
for the first time, we have demonstrated that the overshoot gas from the 
far dust lane coexists with the near dust lane at an similar inclination 
angle to that of the Galactic bar
\citep[$\phi_{\rm bar}=23^{\circ}\pm3^{\circ}$, see details in][]{2024ApJ...971L...6S}. 
The overshoot effect is critical in regulating the secular dynamical 
evolution of gas cycles in the inner Galaxy.\\

2. The overshoot clouds are decelerating (G7), reaccelerating (G5), and 
eventually converging near the bar channel (i.e., DL1 with a cross-sectional 
diameter of $\sim$~320~pc). Undoubtedly, a substantial amount of molecular gas 
is constantly changing its state of motion and spiraling toward the CMZ 
due to the gravitational potential of the bar. 
The total mass of the gas with lower angular momentum, and related to 
the overshoot effect, is about 2 times that of the fresh gas accreted directly 
from the Galactic disk at $R_{\rm GC}\sim$~3--4~kpc (Table~1). This indicates 
an average inflow efficiency of 1/3 for the current period. \\

\begin{figure}
\vspace{-15ex}
\gridline{\hspace{-5ex}  \fig{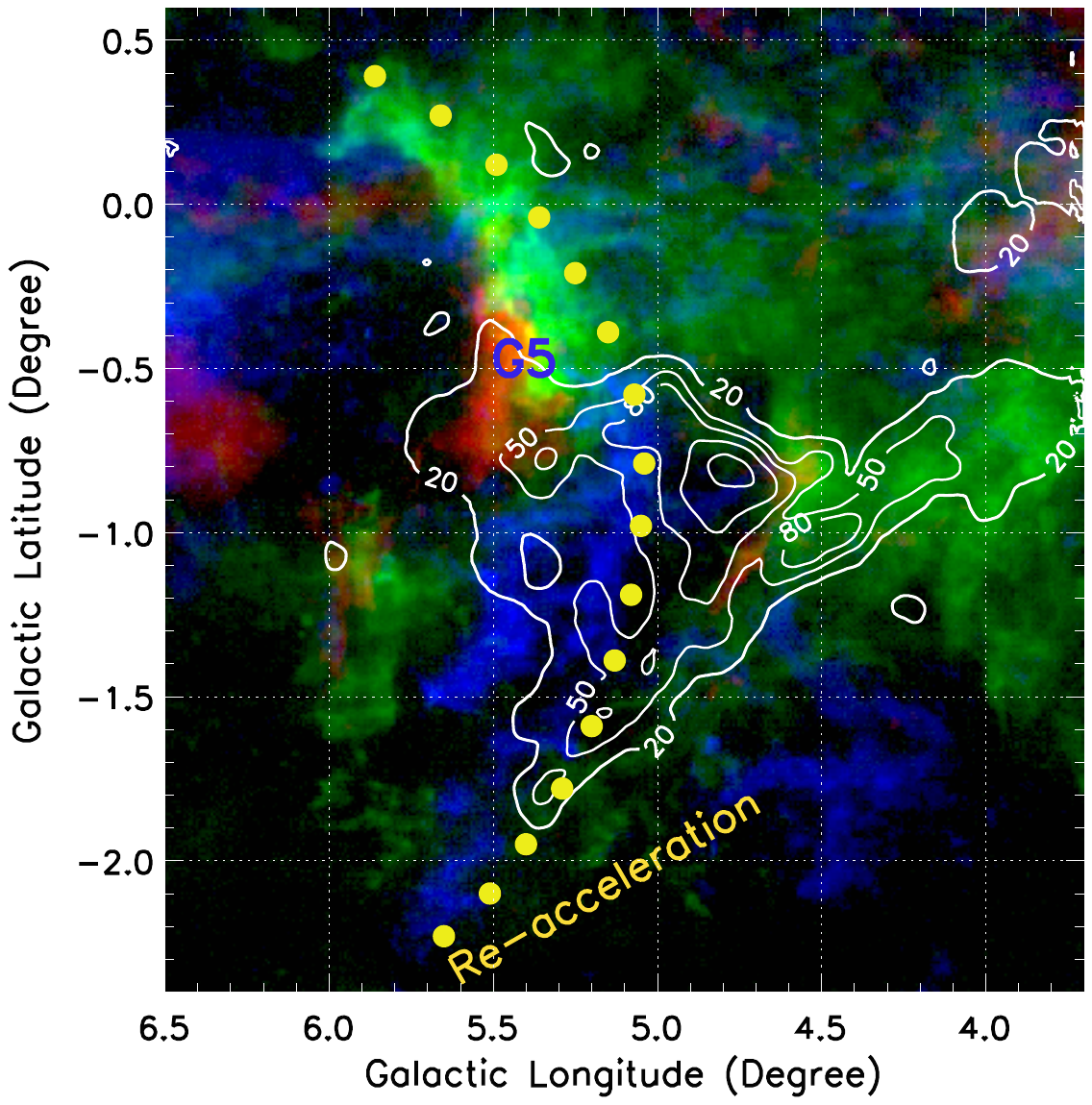}{0.55\textwidth}
           {\vspace{-81.5ex} \hspace{+7.ex}  (a) G5 region and the associated bow-like MCs.}
         \hspace{-7.ex}  \fig{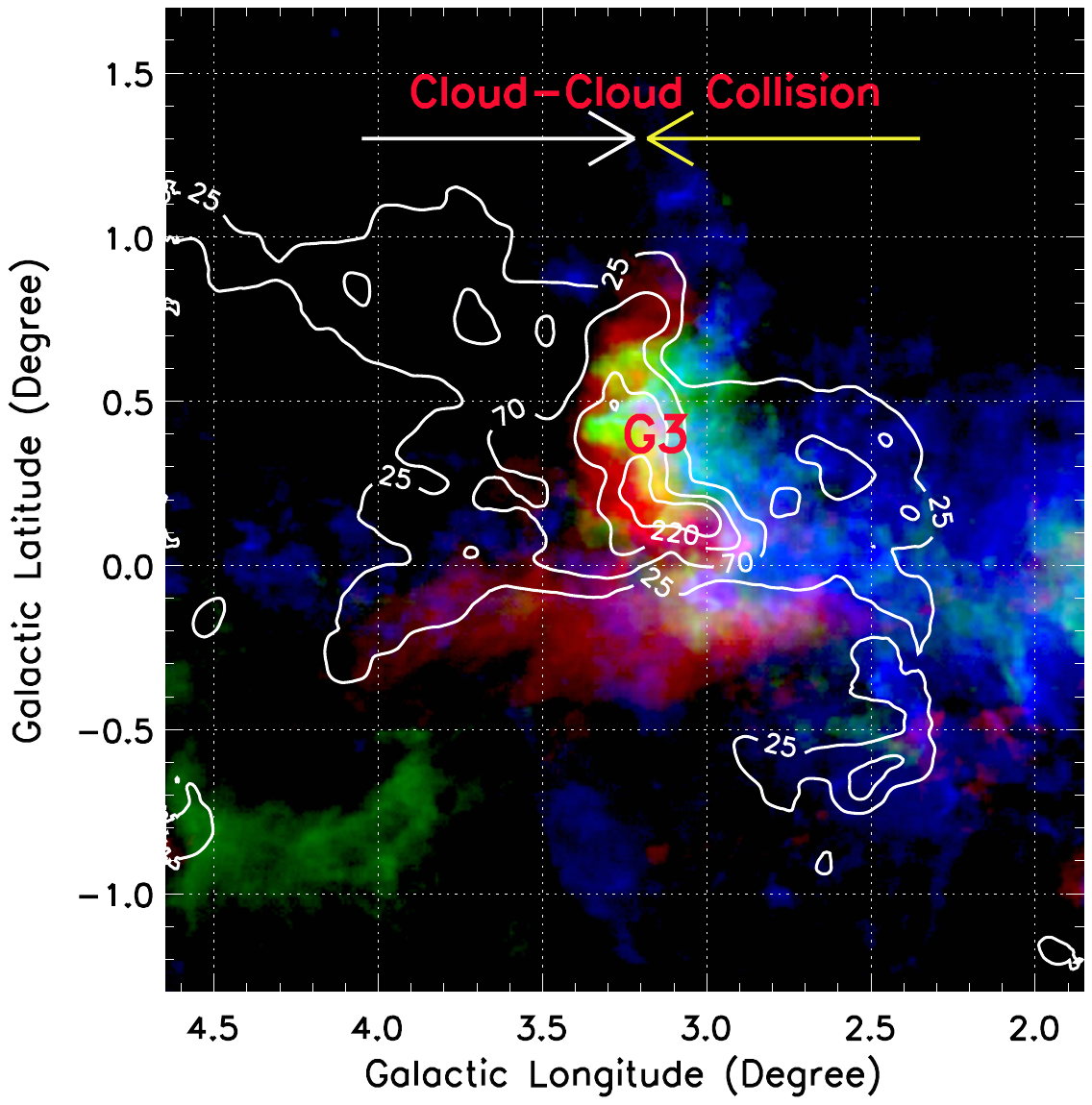}{0.55\textwidth}
           {\vspace{-81.5ex} \hspace{+6.5ex}  (b) G3 region related to the cloud--cloud collision.}
         }
\vspace{-40.ex}
\gridline{\hspace{-7.ex} \fig{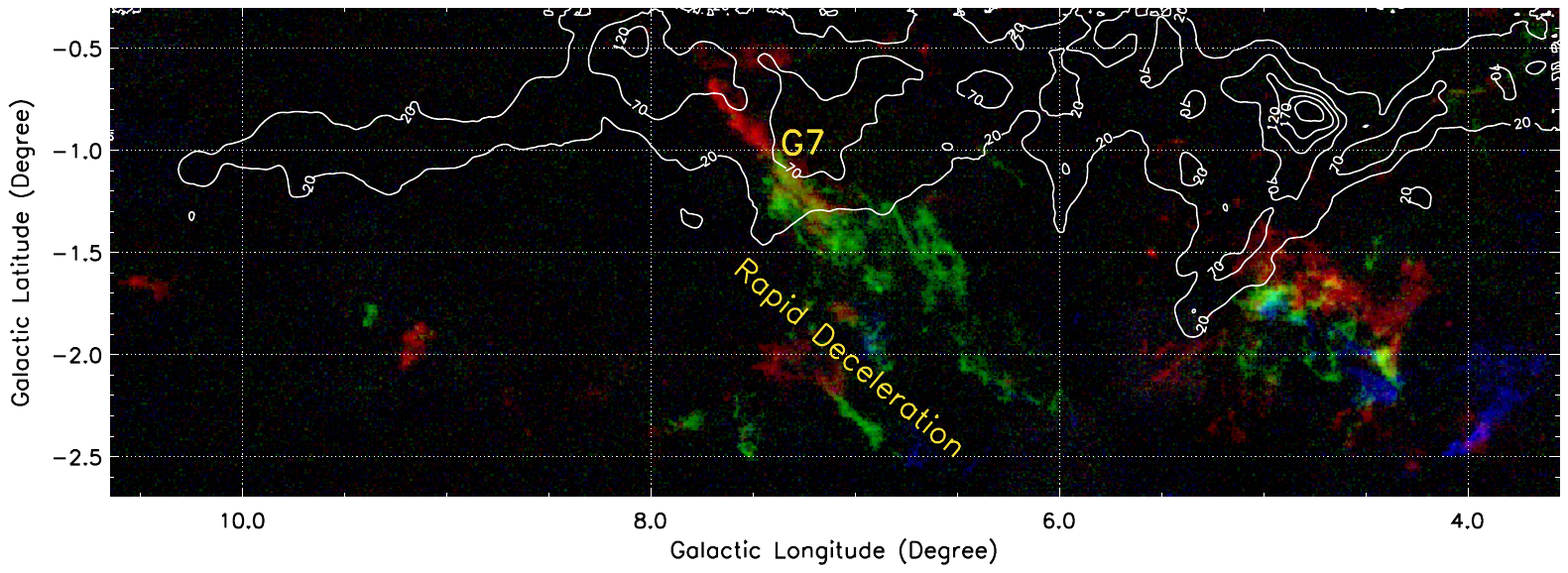}{1.112\textwidth}
{\vspace{-74.ex} \hspace{5ex} (c) Overshoot gas and the associated gas in the near dust lane.}
         }
\vspace{-28ex}
\caption{
Three cases for the physical association between the molecular gas 
in the near dust lane and the overshoot gas from the far dust lane. 
Panel (a) illustrates bow-like MCs that are spatially coherent over
a wide range of velocities ($[-25, -5] \km\ps$ for blue, $[+27, +80] \km\ps$ 
for green, $[+120, +170] \km\ps$ for red, and $[+180, +240] \km\ps$ 
for white contours). The large-scale shock structure traced by
CO emission is indicated by a yellow dotted line.
The G5 region at ($l\sim5\fdg4$, $b\sim-0\fdg4$) or $R_{\rm GC}\sim$1.6~kpc 
is labeled on the figure.
Panel (b) shows the cloud--cloud collision between the approaching gas 
from the overshoot effect ($[-20, +5] \km\ps$ for blue, $[+25, +60] \km\ps$ for green,
and $[+80, +120] \km\ps$ for red) and the receding gas
along the DL2 ($[+130, +180] \km\ps$ for white contours).
The G3 region at ($l\sim3\fdg2$, $b\sim0\fdg3$) or $R_{\rm GC}\sim$0.5~kpc 
is labeled on the figure.
Panel (c) shows the spatial correlation between the overshoot gas
($[-105, -70] \km\ps$ for blue, $[-70, -55] \km\ps$ for green,
and $[-55, -40] \km\ps$ for red) and the DL1 \citep[white contours for 
the near dust lane; also refer to Figure 1 in][]{2024ApJ...971L...6S}.
The G7 region at ($l\sim7\fdg3$, $b\sim-1\fdg1$) or $R_{\rm GC}\sim$2.2~kpc 
is labeled on the figure.
White contours in the three panels mark different levels of the 
integrated CO intensity in K~km~s$^{-1}$. 
\label{fig:f2}}
\end{figure}

3. The cloud--cloud collision features between the overshoot gas from the 
far dust lane and the gas inflows toward the CMZ are clearly evident 
in the G3 and G5 regions (Figure~\ref{fig:f2}; see details in Section 2). 
Other interaction features between gas flows and the Galactic bar are 
also well illustrated, such as changes in gas distribution and kinematics, 
shearing/nonrelaxed gas features relative to the bar, frequent collisions 
and convergence between clouds with different trajectories, and distorted 
bar lanes, etc.  \\

4. As one of the most important gas components, molecular gas undergoes
nonnegligible DISSIPATION processes due to strong turbulence. During 
various dynamical processes, including frequent inelastic collisions, 
viscosity between gas flows, and shear from the bar's torque, 
the angular momentum of the gas is efficiently reduced 
(i.e, DL2+DL3 vs. DL1 in Section 3). These processes collectively 
facilitate the inward migration of molecular gas in the x1 orbit, 
ultimately settling the gas with lower angular momentum into 
the x2 orbit (i.e., the CMZ) over a few tens of Myr. \\

5. The overshoot flows exhibit a more pronounced tilted feature compared
to the near dust lane. We suggest that the twisted $\infty$ structure in the
CMZ \citep[][]{2011ApJ...735L..33M} originates from the combined effects of
the tilted gas flows on kiloparsec scales and the local warped gas structure
dominated by the strong gravitational potential of the nuclear stellar disk.
The nonaxisymmetric gravitational potential of the Galactic bar
serves as an effective regulator of angular momentum redistribution, mass transport, 
and the secular evolution of the inner Galaxy (see Section 4).  \\

6. The gas inflow rate toward the CMZ is $\gsim$1.1~$\Msun$~yr$^{-1}$
according to the total gas input from the fresh gas of the near dust lane 
and the overshoot gas from the far dust lane. The average gas inflow rate toward
the CMZ is roughly comparable to the outflow's rate for 
the Galactic nuclear winds \citep[][]{2022ApJ...930..112S}. 
However, the gas inflow rate exhibits periodic variations of several Myr 
due to the overshoot effect \citep[also refer to][]{2021ApJ...922...79H} 
and significant instantaneous fluctuations (e.g., $\sim$5--10 
times higher than the average gas inflow rate) due to gas 
accumulation in some regions (e.g., G3 at 
$R_{\rm GC}\sim\frac{1}{7}R_{\rm bar}$, G5 at $R_{\rm GC}\sim\frac{1}{2}R_{\rm bar}$,
and G7 at $R_{\rm GC}\sim\frac{2}{3}R_{\rm bar}$).  \\

\begin{figure}
\vspace{-33.5ex}
\gridline{\hspace{-0.5ex} \fig{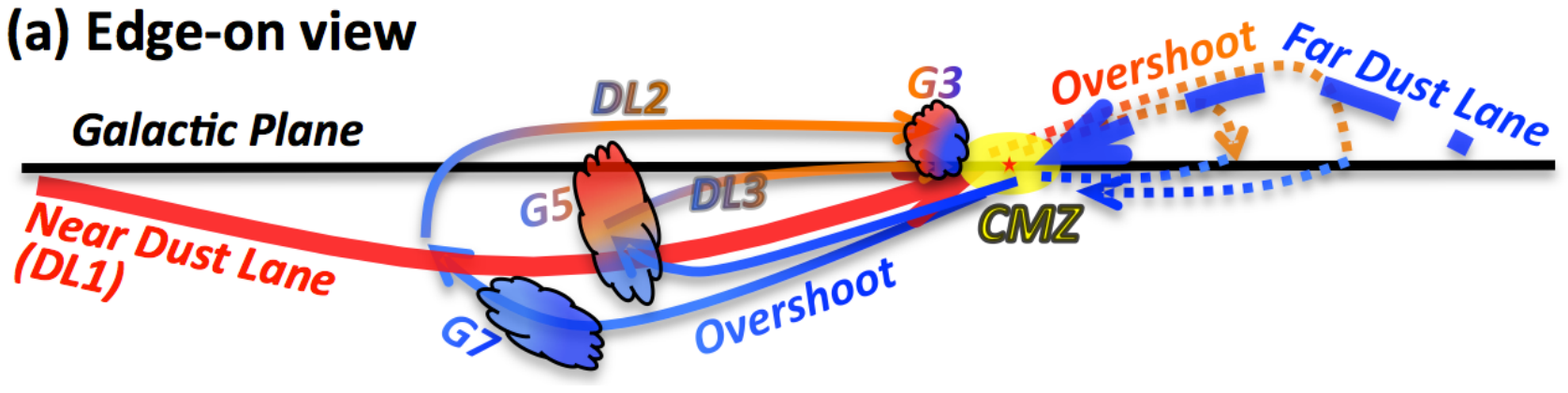}{1.\textwidth}
          {\vspace{-10.ex} \hspace{1ex}}   
         }
\vspace{-42.ex}
\gridline{\hspace{+1.ex}  \fig{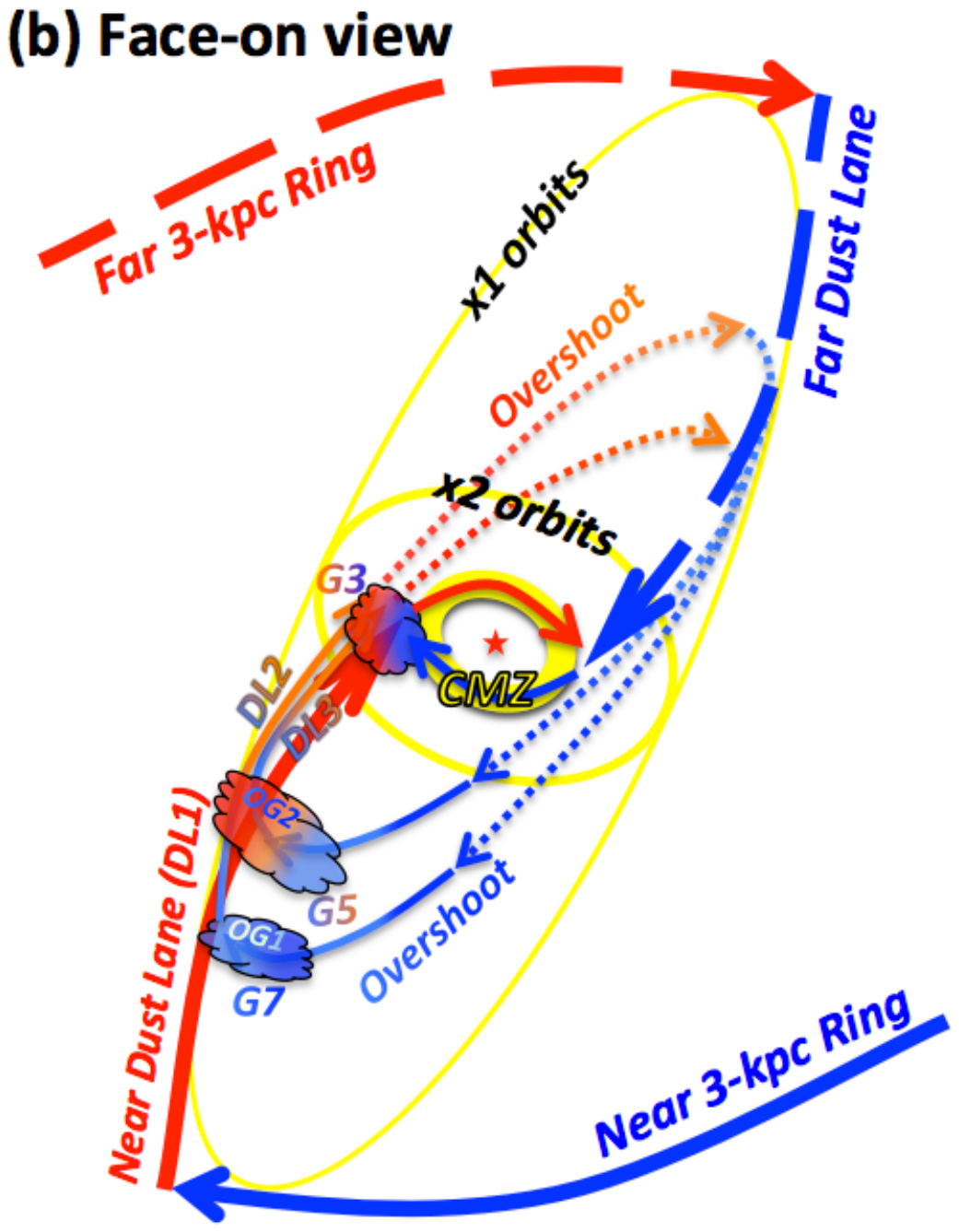}{0.65\textwidth}
           {\vspace{-100.ex} \hspace{+8.5ex}} 
         }
\vspace{-18ex}
\caption{
Holistic perspective of the large-scale molecular gas flows toward 
the CMZ based on the MWISP CO data and other supplementary data.
In the diagram, the upper portion displays an edge-on view
(from the Sun) of the identified gas structures and some prominent 
features by considering the observation effect 
\citep[e.g., see Figure 2 in][]{2023ApJ...959...93G}, while the lower 
portion shows a face-on view of the corresponding structures.
Note that the $b$ scale (the Galactic latitude) has been enlarged 
to better show the twisted or warped structure of the gas flows.
The parameters of the geometry are from \cite{2024ApJ...971L...6S},
i.e., the inclination angle of $\phi_{\rm bar}=23^{\circ}\pm3^{\circ}$,
the bar length of $\sim3.2-3.4$~kpc, and the tilted angle of 
$\theta_{\rm{bar\ lanes}}\gsim 5^{\circ}$ for the large-scale
gas feature from $l\sim+6^{\circ}$ to $-4^{\circ}$.
Here we adopt $R_{0}$=8.15~kpc \citep[][]{Reid19}. Solid lines 
represent the near-side gas flows from the MWISP observations, 
while dotted lines show the far-side gas flows based on other CO 
surveys and the mirrored MWISP data.
Dominated by the nonaxisymmetric gravitational potential of the 
Galactic bar, the overshoot effect plays a crucial role in
redistributing angular momentum outward through various dynamical 
processes such as shocks generated by bar perturbations, frequent 
inelastic collisions, and nonnegligible viscosity.
G5 and G7 are located near the $R_{\rm GC}\sim\frac{1}{2}R_{\rm bar}$
and $R_{\rm GC}\sim\frac{2}{3}R_{\rm bar}$ regions, respectively. 
OG1 and OG2 (see Figure~\ref{fig:f1}) in the map roughly show the trajectory of 
the gas flows from approaching to receding. As a result, 
a large amount of molecular gas located several kiloparsecs away can efficiently 
migrate into regions within a few hundred parsecs of the GC over a few 
tens of millions of years, leading to a burst of star formation in the 
CMZ and facilitating the growth of the central SMBH.
\label{fig:f3}}
\end{figure}

7. The MWISP survey with the high-sensitivity and high-velocity 
resolution can dramatically enhance our ability to capture the 
detailed morphology and kinematics of gas flows toward the CMZ, 
helping us to reconstruct the entire dynamical evolution of 
gas flows within $\sim$3--4~kpc of the GC (Figure~\ref{fig:f3}). 
Large-scale gas accretion and overshoot effects, combined with various 
feedback within a few hundred parsecs of the GC, jointly shape the gas 
distribution, physical properties, and dynamical evolution of the CMZ, 
as well as the subsequent star formation processes therein.  \\

The holistic structure and large-sample statistical properties 
revealed by the observations are important for our understanding 
of the gas dynamics toward the center of galaxies. Moreover, 
many simulation studies have been done to explain how gas structures
form within a bar and how they depend on various bar parameters \citep[e.g., 
the bar strength, the effective sound speed, the pattern speed of the bar, 
the inclination angle of the bar, the magnetic field, 
the various gravitational potentials, etc.; see][]{1996A&A...313...65L,
2012ApJ...758...14K,2012ApJ...747...60K,2012ApJ...751..124K,2024ApJ...963...22F,
2024MNRAS.528.5742S,2024A&A...691A.303T}.
The comprehensive analysis of large-scale survey data, combined
with the comparison of observational features and model simulations,
is essential for accurately understanding the astrophysical phenomena
and fundamental physical processes that dominate galaxy evolution.

\acknowledgments
We would like to thank the referee for several constructive comments 
that led to improvements in the paper.
This research made use of the data from the Milky Way Imaging Scroll Painting 
(MWISP) project, which is a multiline survey in \twCO/\thCO/C$^{18}$O along the 
northern Galactic plane with the PMO-13.7m telescope. We are grateful to all the members 
of the MWISP working group, particularly the staff members at the PMO-13.7m telescope, 
for their long-term support. MWISP was sponsored by National Key R\&D Program of 
China with grants 2023YFA1608000 and 2017YFA0402701 and by CAS Key Research Program 
of Frontier Sciences with grant QYZDJ-SSW-SLH047.
We also acknowledge support from the National Natural Science Foundation of China 
through grants 12173090 and 12041305.

\facility{PMO:DLH}

\bibliography{references}{}

\begin{thebibliography}{}
\expandafter\ifx\csname natexlab\endcsname\relax\def\natexlab#1{#1}\fi
\providecommand{\url}[1]{\href{#1}{#1}}
\providecommand{\dodoi}[1]{doi:~\href{http://doi.org/#1}{\nolinkurl{#1}}}
\providecommand{\doeprint}[1]{\href{http://ascl.net/#1}{\nolinkurl{http://ascl.net/#1}}}
\providecommand{\doarXiv}[1]{\href{https://arxiv.org/abs/#1}{\nolinkurl{https://arxiv.org/abs/#1}}}

\bibitem[{{Akhter} {et~al.}(2021){Akhter}, {Cunningham}, {Harvey-Smith},
  {Nawaz}, {Jones}, {Walsh}, {de Gouveia Dal Pino}, \&
  {Falceta-Gon{\c{c}}alves}}]{2021MNRAS.502.5896A}
{Akhter}, S., {Cunningham}, M.~R., {Harvey-Smith}, L., {et~al.} 2021, \mnras,
  502, 5896, \dodoi{10.1093/mnras/staa267}

\bibitem[{{Anderson} {et~al.}(2020){Anderson}, {Sormani}, {Ginsburg}, {Glover},
  {Heywood}, {Rammala}, {Schuller}, {Csengeri}, {Urquhart}, \&
  {Bronfman}}]{2020ApJ...901...51A}
{Anderson}, L.~D., {Sormani}, M.~C., {Ginsburg}, A., {et~al.} 2020, \apj, 901,
  51, \dodoi{10.3847/1538-4357/abadf6}

\bibitem[{{Armillotta} {et~al.}(2019){Armillotta}, {Krumholz}, {Di Teodoro}, \&
  {McClure-Griffiths}}]{2019MNRAS.490.4401A}
{Armillotta}, L., {Krumholz}, M.~R., {Di Teodoro}, E.~M., \&
  {McClure-Griffiths}, N.~M. 2019, \mnras, 490, 4401,
  \dodoi{10.1093/mnras/stz2880}

\bibitem[{{Athanassoula}(1992)}]{1992MNRAS.259..345A}
{Athanassoula}, E. 1992, \mnras, 259, 345, \dodoi{10.1093/mnras/259.2.345}

\bibitem[{{Battersby} {et~al.}(2024){Battersby}, {Walker}, {Barnes},
  {Ginsburg}, {Lipman}, {Alboslani}, {Hatchfield}, {Bally}, {Glover},
  {Henshaw}, {Immer}, {Klessen}, {Longmore}, {Mills}, {Molinari}, {Smith},
  {Sormani}, {Tress}, \& {Zhang}}]{2024arXiv241017334B}
{Battersby}, C., {Walker}, D.~L., {Barnes}, A., {et~al.} 2024, arXiv e-prints,
  arXiv:2410.17334.
\newblock \doarXiv{2410.17334}

\bibitem[{{Bland-Hawthorn} \& {Gerhard}(2016)}]{2016ARA&A..54..529B}
{Bland-Hawthorn}, J., \& {Gerhard}, O. 2016, \araa, 54, 529,
  \dodoi{10.1146/annurev-astro-081915-023441}

\bibitem[{{Bolatto} {et~al.}(2013){Bolatto}, {Wolfire}, \&
  {Leroy}}]{2013ARA&A..51..207B}
{Bolatto}, A.~D., {Wolfire}, M., \& {Leroy}, A.~K. 2013, \araa, 51, 207,
  \dodoi{10.1146/annurev-astro-082812-140944}

\bibitem[{{Bovy} {et~al.}(2019){Bovy}, {Leung}, {Hunt}, {Mackereth},
  {Garc{\'\i}a-Hern{\'a}ndez}, \& {Roman-Lopes}}]{2019MNRAS.490.4740B}
{Bovy}, J., {Leung}, H.~W., {Hunt}, J. A.~S., {et~al.} 2019, \mnras, 490, 4740,
  \dodoi{10.1093/mnras/stz2891}

\bibitem[{{Bryant} \& {Krabbe}(2021)}]{2021NewAR..9301630B}
{Bryant}, A., \& {Krabbe}, A. 2021, \nar, 93, 101630,
  \dodoi{10.1016/j.newar.2021.101630}

\bibitem[{{Burton} \& {Liszt}(1983)}]{1983A&AS...52...63B}
{Burton}, W.~B., \& {Liszt}, H.~S. 1983, \aaps, 52, 63

\bibitem[{{Busch} {et~al.}(2022){Busch}, {Riquelme}, {G{\"u}sten}, {Menten},
  {Pillai}, \& {Kauffmann}}]{2022A&A...668A.183B}
{Busch}, L.~A., {Riquelme}, D., {G{\"u}sten}, R., {et~al.} 2022, \aap, 668,
  A183, \dodoi{10.1051/0004-6361/202244870}

\bibitem[{{Butterfield} {et~al.}(2025){Butterfield}, {Morgan}, {Barnes},
  {Ginsburg}, {Gramze}, {Morris}, {Sormani}, {Battersby}, {Burton}, {Costa},
  {Mills}, {Ott}, \& {Rugel}}]{2025arXiv250314174B}
{Butterfield}, N., {Morgan}, L., {Barnes}, A., {et~al.} 2025, arXiv e-prints,
  arXiv:2503.14174.
\newblock \doarXiv{2503.14174}

\bibitem[{{Chaves-Velasquez} {et~al.}(2025){Chaves-Velasquez}, {G{\'o}mez}, \&
  {P{\'e}rez-Villegas}}]{2025PASA...42...14C}
{Chaves-Velasquez}, L., {G{\'o}mez}, G.~C., \& {P{\'e}rez-Villegas}, {\'A}.
  2025, \pasa, 42, e014, \dodoi{10.1017/pasa.2024.130}

\bibitem[{{Chiba} {et~al.}(2021){Chiba}, {Friske}, \&
  {Sch{\"o}nrich}}]{2021MNRAS.500.4710C}
{Chiba}, R., {Friske}, J. K.~S., \& {Sch{\"o}nrich}, R. 2021, \mnras, 500,
  4710, \dodoi{10.1093/mnras/staa3585}

\bibitem[{{Clarke} \& {Gerhard}(2022)}]{2022MNRAS.512.2171C}
{Clarke}, J.~P., \& {Gerhard}, O. 2022, \mnras, 512, 2171,
  \dodoi{10.1093/mnras/stac603}

\bibitem[{{Combes} {et~al.}(2014){Combes}, {Garc{\'\i}a-Burillo}, {Casasola},
  {Hunt}, {Krips}, {Baker}, {Boone}, {Eckart}, {Marquez}, {Neri}, {Schinnerer},
  \& {Tacconi}}]{2014A&A...565A..97C}
{Combes}, F., {Garc{\'\i}a-Burillo}, S., {Casasola}, V., {et~al.} 2014, \aap,
  565, A97, \dodoi{10.1051/0004-6361/201423433}

\bibitem[{{Combes} {et~al.}(2019){Combes}, {Garc{\'\i}a-Burillo}, {Audibert},
  {Hunt}, {Eckart}, {Aalto}, {Casasola}, {Boone}, {Krips}, {Viti}, {Sakamoto},
  {Muller}, {Dasyra}, {van der Werf}, \& {Martin}}]{2019A&A...623A..79C}
{Combes}, F., {Garc{\'\i}a-Burillo}, S., {Audibert}, A., {et~al.} 2019, \aap,
  623, A79, \dodoi{10.1051/0004-6361/201834560}

\bibitem[{{Dahmen} {et~al.}(1998){Dahmen}, {Huttemeister}, {Wilson}, \&
  {Mauersberger}}]{1998A&A...331..959D}
{Dahmen}, G., {Huttemeister}, S., {Wilson}, T.~L., \& {Mauersberger}, R. 1998,
  \aap, 331, 959, \dodoi{10.48550/arXiv.astro-ph/9711117}

\bibitem[{{Dame} {et~al.}(2001){Dame}, {Hartmann}, \&
  {Thaddeus}}]{2001ApJ...547..792D}
{Dame}, T.~M., {Hartmann}, D., \& {Thaddeus}, P. 2001, \apj, 547, 792,
  \dodoi{10.1086/318388}

\bibitem[{{Dame} \& {Thaddeus}(2008)}]{2008Dame}
{Dame}, T.~M., \& {Thaddeus}, P. 2008, \apjl, 683, L143, \dodoi{10.1086/591669}

\bibitem[{{Dillamore} {et~al.}(2023){Dillamore}, {Belokurov}, {Evans}, \&
  {Davies}}]{2023MNRAS.524.3596D}
{Dillamore}, A.~M., {Belokurov}, V., {Evans}, N.~W., \& {Davies}, E.~Y. 2023,
  \mnras, 524, 3596, \dodoi{10.1093/mnras/stad2136}

\bibitem[{{Eden} {et~al.}(2020){Eden}, {Moore}, {Currie}, {Rigby},
  {Rosolowsky}, {Su}, {Kim}, {Parsons}, {Morata}, {Chen}, {Minamidani}, {Park},
  {Ragan}, {Urquhart}, {Rani}, {Tahani}, {Billington}, {Deb}, {Figura},
  {Fujiyoshi}, {Joncas}, {Liao}, {Liu}, {Ma}, {Tuan-Anh}, {Yun}, {Zhang},
  {Zhu}, {Henshaw}, {Longmore}, {Kobayashi}, {Thompson}, {Ao},
  {Campbell-White}, {Ching}, {Chung}, {Duarte-Cabral}, {Fich}, {Gao}, {Graves},
  {Jiang}, {Kemper}, {Kuan}, {Kwon}, {Lee}, {Lee}, {Liu}, {Pe{\~n}aloza},
  {Peretto}, {Phuong}, {Pineda}, {Plume}, {Puspitaningrum}, {Samal}, {Soam},
  {Sun}, {Tang}, {Traficante}, {White}, {Yan}, {Yang}, {Yuan}, {Yue}, {Bemis},
  {Brunt}, {Chen}, {Cho}, {Clark}, {Cyganowski}, {Friberg}, {Fuller}, {Han},
  {Hoare}, {Izumi}, {Kim}, {Kim}, {Kim}, {Koch}, {Kuno}, {Lacialle}, {Lai},
  {Lee}, {Lee}, {Li}, {Liu}, {Mairs}, {Pan}, {Qian}, {Scicluna}, {Shi}, {Shi},
  {Srinivasan}, {Tan}, {Thomas}, {Torii}, {Trejo}, {Umemoto}, {Violino},
  {Wallstr{\"o}m}, {Wang}, {Wu}, {Yuan}, {Zhang}, {Zhang}, {Zhou}, \&
  {Zhou}}]{2020MNRAS.498.5936E}
{Eden}, D.~J., {Moore}, T.~J.~T., {Currie}, M.~J., {et~al.} 2020, \mnras, 498,
  5936, \dodoi{10.1093/mnras/staa2734}

\bibitem[{{Emsellem} {et~al.}(2015){Emsellem}, {Renaud}, {Bournaud},
  {Elmegreen}, {Combes}, \& {Gabor}}]{2015MNRAS.446.2468E}
{Emsellem}, E., {Renaud}, F., {Bournaud}, F., {et~al.} 2015, \mnras, 446, 2468,
  \dodoi{10.1093/mnras/stu2209}

\bibitem[{{Englmaier} \& {Gerhard}(1999)}]{1999MNRAS.304..512E}
{Englmaier}, P., \& {Gerhard}, O. 1999, \mnras, 304, 512,
  \dodoi{10.1046/j.1365-8711.1999.02280.x}

\bibitem[{{Enokiya} {et~al.}(2021){Enokiya}, {Torii}, \&
  {Fukui}}]{2021PASJ...73S..75E}
{Enokiya}, R., {Torii}, K., \& {Fukui}, Y. 2021, \pasj, 73, S75,
  \dodoi{10.1093/pasj/psz119}

\bibitem[{{Feng} {et~al.}(2024){Feng}, {Li}, {Shen}, {Gerhard}, {Saglia},
  {Bla{\~n}a}, {Li}, \& {Jing}}]{2024ApJ...963...22F}
{Feng}, Z.-X., {Li}, Z., {Shen}, J., {et~al.} 2024, \apj, 963, 22,
  \dodoi{10.3847/1538-4357/ad13ee}

\bibitem[{{Ferri{\`e}re} {et~al.}(2007){Ferri{\`e}re}, {Gillard}, \&
  {Jean}}]{2007A&A...467..611F}
{Ferri{\`e}re}, K., {Gillard}, W., \& {Jean}, P. 2007, \aap, 467, 611,
  \dodoi{10.1051/0004-6361:20066992}

\bibitem[{{Fukui} {et~al.}(2021){Fukui}, {Habe}, {Inoue}, {Enokiya}, \&
  {Tachihara}}]{2021PASJ...73S...1F}
{Fukui}, Y., {Habe}, A., {Inoue}, T., {Enokiya}, R., \& {Tachihara}, K. 2021,
  \pasj, 73, S1, \dodoi{10.1093/pasj/psaa103}

\bibitem[{{Furukawa} {et~al.}(2009){Furukawa}, {Dawson}, {Ohama}, {Kawamura},
  {Mizuno}, {Onishi}, \& {Fukui}}]{2009ApJ...696L.115F}
{Furukawa}, N., {Dawson}, J.~R., {Ohama}, A., {et~al.} 2009, \apjl, 696, L115,
  \dodoi{10.1088/0004-637X/696/2/L115}

\bibitem[{{Fux}(1999)}]{1999A&A...345..787F}
{Fux}, R. 1999, \aap, 345, 787, \dodoi{10.48550/arXiv.astro-ph/9903154}

\bibitem[{{Gong} {et~al.}(2017){Gong}, {Fang}, {Mao}, {Zhang}, {Wang}, {Su},
  {Chen}, {Yang}, {Wang}, \& {Lu}}]{2017ApJ...835L..14G}
{Gong}, Y., {Fang}, M., {Mao}, R., {et~al.} 2017, \apjl, 835, L14,
  \dodoi{10.3847/2041-8213/835/1/L14}

\bibitem[{{Gramze} {et~al.}(2023){Gramze}, {Ginsburg}, {Meier}, {Ott},
  {Shirley}, {Sormani}, \& {Svoboda}}]{2023ApJ...959...93G}
{Gramze}, S.~R., {Ginsburg}, A., {Meier}, D.~S., {et~al.} 2023, \apj, 959, 93,
  \dodoi{10.3847/1538-4357/ad01be}

\bibitem[{{Hatchfield} {et~al.}(2021){Hatchfield}, {Sormani}, {Tress},
  {Battersby}, {Smith}, {Glover}, \& {Klessen}}]{2021ApJ...922...79H}
{Hatchfield}, H.~P., {Sormani}, M.~C., {Tress}, R.~G., {et~al.} 2021, \apj,
  922, 79, \dodoi{10.3847/1538-4357/ac1e89}

\bibitem[{{Henshaw} {et~al.}(2023){Henshaw}, {Barnes}, {Battersby}, {Ginsburg},
  {Sormani}, \& {Walker}}]{2023ASPC..534...83H}
{Henshaw}, J.~D., {Barnes}, A.~T., {Battersby}, C., {et~al.} 2023, in
  Astronomical Society of the Pacific Conference Series, Vol. 534, Protostars
  and Planets VII, ed. S.~{Inutsuka}, Y.~{Aikawa}, T.~{Muto}, K.~{Tomida}, \&
  M.~{Tamura}, 83, \dodoi{10.48550/arXiv.2203.11223}

\bibitem[{{Henshaw} {et~al.}(2019){Henshaw}, {Ginsburg}, {Haworth}, {Longmore},
  {Kruijssen}, {Mills}, {Sokolov}, {Walker}, {Barnes}, {Contreras}, {Bally},
  {Battersby}, {Beuther}, {Butterfield}, {Dale}, {Henning}, {Jackson},
  {Kauffmann}, {Pillai}, {Ragan}, {Riener}, \& {Zhang}}]{2019MNRAS.485.2457H}
{Henshaw}, J.~D., {Ginsburg}, A., {Haworth}, T.~J., {et~al.} 2019, \mnras, 485,
  2457, \dodoi{10.1093/mnras/stz471}

\bibitem[{{Heywood} {et~al.}(2019){Heywood}, {Camilo}, {Cotton}, {Yusef-Zadeh},
  {Abbott}, {Adam}, {Aldera}, {Bauermeister}, {Booth}, {Botha}, {Botha},
  {Brederode}, {Brits}, {Buchner}, {Burger}, {Chalmers}, {Cheetham}, {de
  Villiers}, {Dikgale-Mahlakoana}, {du Toit}, {Esterhuyse}, {Fanaroff},
  {Foley}, {Fourie}, {Gamatham}, {Goedhart}, {Gounden}, {Hlakola}, {Hoek},
  {Hokwana}, {Horn}, {Horrell}, {Hugo}, {Isaacson}, {Jonas}, {Jordaan},
  {Joubert}, {J{\'o}zsa}, {Julie}, {Kapp}, {Kenyon}, {Kotz{\'e}}, {Kriel},
  {Kusel}, {Lehmensiek}, {Liebenberg}, {Loots}, {Lord}, {Lunsky}, {Macfarlane},
  {Magnus}, {Magozore}, {Mahgoub}, {Main}, {Malan}, {Malgas}, {Manley},
  {Maree}, {Merry}, {Millenaar}, {Mnyandu}, {Moeng}, {Monama}, {Mphego}, {New},
  {Ngcebetsha}, {Oozeer}, {Otto}, {Passmoor}, {Patel}, {Peens-Hough},
  {Perkins}, {Ratcliffe}, {Renil}, {Rust}, {Salie}, {Schwardt}, {Serylak},
  {Siebrits}, {Sirothia}, {Smirnov}, {Sofeya}, {Swart}, {Tasse}, {Taylor},
  {Theron}, {Thorat}, {Tiplady}, {Tshongweni}, {van Balla}, {van der Byl}, {van
  der Merwe}, {van Dyk}, {Van Rooyen}, {Van Tonder}, {Van Wyk}, {Wallace},
  {Welz}, \& {Williams}}]{2019Natur.573..235H}
{Heywood}, I., {Camilo}, F., {Cotton}, W.~D., {et~al.} 2019, \nat, 573, 235,
  \dodoi{10.1038/s41586-019-1532-5}

\bibitem[{{Hilmi} {et~al.}(2020){Hilmi}, {Minchev}, {Buck}, {Martig},
  {Quillen}, {Monari}, {Famaey}, {de Jong}, {Laporte}, {Read}, {Sanders},
  {Steinmetz}, \& {Wegg}}]{2020MNRAS.497..933H}
{Hilmi}, T., {Minchev}, I., {Buck}, T., {et~al.} 2020, \mnras, 497, 933,
  \dodoi{10.1093/mnras/staa1934}

\bibitem[{{Inoue} \& {Fukui}(2013)}]{2013ApJ...774L..31I}
{Inoue}, T., \& {Fukui}, Y. 2013, \apjl, 774, L31,
  \dodoi{10.1088/2041-8205/774/2/L31}

\bibitem[{{Karoly} {et~al.}(2025){Karoly}, {Ward-Thompson}, {Pattle},
  {Longmore}, {Di Francesco}, {Whitworth}, {Johnstone}, {Sadavoy}, {Koch},
  {Yang}, {Furuya}, {Lu}, {Tamura}, {Debattista}, {Eden}, {Hwang}, {Poidevin},
  {Bijas}, {Chen}, {Chung}, {Coud{\'e}}, {Lin}, {Doi}, {Onaka}, {Fanciullo},
  {Liu}, {Li}, {Bastien}, {Hasegawa}, {Kwon}, {Lai}, \&
  {Qiu}}]{2025ApJ...982L..22K}
{Karoly}, J., {Ward-Thompson}, D., {Pattle}, K., {et~al.} 2025, \apjl, 982,
  L22, \dodoi{10.3847/2041-8213/adbc67}

\bibitem[{{Kim} {et~al.}(2012{\natexlab{a}}){Kim}, {Seo}, \&
  {Kim}}]{2012ApJ...758...14K}
{Kim}, W.-T., {Seo}, W.-Y., \& {Kim}, Y. 2012{\natexlab{a}}, \apj, 758, 14,
  \dodoi{10.1088/0004-637X/758/1/14}

\bibitem[{{Kim} {et~al.}(2012{\natexlab{b}}){Kim}, {Seo}, {Stone}, {Yoon}, \&
  {Teuben}}]{2012ApJ...747...60K}
{Kim}, W.-T., {Seo}, W.-Y., {Stone}, J.~M., {Yoon}, D., \& {Teuben}, P.~J.
  2012{\natexlab{b}}, \apj, 747, 60, \dodoi{10.1088/0004-637X/747/1/60}

\bibitem[{{Kim} \& {Stone}(2012)}]{2012ApJ...751..124K}
{Kim}, W.-T., \& {Stone}, J.~M. 2012, \apj, 751, 124,
  \dodoi{10.1088/0004-637X/751/2/124}

\bibitem[{{Koda} {et~al.}(2016){Koda}, {Scoville}, \&
  {Heyer}}]{2016ApJ...823...76K}
{Koda}, J., {Scoville}, N., \& {Heyer}, M. 2016, \apj, 823, 76,
  \dodoi{10.3847/0004-637X/823/2/76}

\bibitem[{{Kohno} \& {Sofue}(2024)}]{2024MNRAS.527.9290K}
{Kohno}, M., \& {Sofue}, Y. 2024, \mnras, 527, 9290,
  \dodoi{10.1093/mnras/stad3648}

\bibitem[{{Kruijssen} {et~al.}(2015){Kruijssen}, {Dale}, \&
  {Longmore}}]{2015MNRAS.447.1059K}
{Kruijssen}, J.~M.~D., {Dale}, J.~E., \& {Longmore}, S.~N. 2015, \mnras, 447,
  1059, \dodoi{10.1093/mnras/stu2526}

\bibitem[{{Kruijssen} {et~al.}(2019){Kruijssen}, {Dale}, {Longmore}, {Walker},
  {Henshaw}, {Jeffreson}, {Petkova}, {Ginsburg}, {Barnes}, {Battersby},
  {Immer}, {Jackson}, {Keto}, {Krieger}, {Mills}, {S{\'a}nchez-Monge},
  {Schmiedeke}, {Suri}, \& {Zhang}}]{2019MNRAS.484.5734K}
{Kruijssen}, J.~M.~D., {Dale}, J.~E., {Longmore}, S.~N., {et~al.} 2019, \mnras,
  484, 5734, \dodoi{10.1093/mnras/stz381}

\bibitem[{{Krumholz} {et~al.}(2018){Krumholz}, {Burkhart}, {Forbes}, \&
  {Crocker}}]{2018MNRAS.477.2716K}
{Krumholz}, M.~R., {Burkhart}, B., {Forbes}, J.~C., \& {Crocker}, R.~M. 2018,
  \mnras, 477, 2716, \dodoi{10.1093/mnras/sty852}

\bibitem[{{Krumholz} {et~al.}(2017){Krumholz}, {Kruijssen}, \&
  {Crocker}}]{2017MNRAS.466.1213K}
{Krumholz}, M.~R., {Kruijssen}, J.~M.~D., \& {Crocker}, R.~M. 2017, \mnras,
  466, 1213, \dodoi{10.1093/mnras/stw3195}

\bibitem[{{Kumar} {et~al.}(2025){Kumar}, {Reid}, {Dame}, {Ellingsen}, {Hyland},
  {Brunthaler}, {Menten}, {Zheng}, \& {Sanna}}]{2025ApJ...982..185K}
{Kumar}, J., {Reid}, M.~J., {Dame}, T.~M., {et~al.} 2025, \apj, 982, 185,
  \dodoi{10.3847/1538-4357/adb70f}

\bibitem[{{Launhardt} {et~al.}(2002){Launhardt}, {Zylka}, \&
  {Mezger}}]{2002A&A...384..112L}
{Launhardt}, R., {Zylka}, R., \& {Mezger}, P.~G. 2002, \aap, 384, 112,
  \dodoi{10.1051/0004-6361:20020017}

\bibitem[{{Leroy} {et~al.}(2021){Leroy}, {Schinnerer}, {Hughes}, {Rosolowsky},
  {Pety}, {Schruba}, {Usero}, {Blanc}, {Chevance}, {Emsellem}, {Faesi},
  {Herrera}, {Liu}, {Meidt}, {Querejeta}, {Saito}, {Sandstrom}, {Sun},
  {Williams}, {Anand}, {Barnes}, {Behrens}, {Belfiore}, {Benincasa},
  {Be{\v{s}}li{\'c}}, {Bigiel}, {Bolatto}, {den Brok}, {Cao}, {Chandar},
  {Chastenet}, {Chiang}, {Congiu}, {Dale}, {Deger}, {Eibensteiner}, {Egorov},
  {Garc{\'\i}a-Rodr{\'\i}guez}, {Glover}, {Grasha}, {Henshaw}, {Ho}, {Kepley},
  {Kim}, {Klessen}, {Kreckel}, {Koch}, {Kruijssen}, {Larson}, {Lee}, {Lopez},
  {Machado}, {Mayker}, {McElroy}, {Murphy}, {Ostriker}, {Pan}, {Pessa},
  {Puschnig}, {Razza}, {S{\'a}nchez-Bl{\'a}zquez}, {Santoro}, {Sardone},
  {Scheuermann}, {Sliwa}, {Sormani}, {Stuber}, {Thilker}, {Turner}, {Utomo},
  {Watkins}, \& {Whitmore}}]{2021ApJS..257...43L}
{Leroy}, A.~K., {Schinnerer}, E., {Hughes}, A., {et~al.} 2021, \apjs, 257, 43,
  \dodoi{10.3847/1538-4365/ac17f3}

\bibitem[{{Li} {et~al.}(2016){Li}, {Gerhard}, {Shen}, {Portail}, \&
  {Wegg}}]{2016ApJ...824...13L}
{Li}, Z., {Gerhard}, O., {Shen}, J., {Portail}, M., \& {Wegg}, C. 2016, \apj,
  824, 13, \dodoi{10.3847/0004-637X/824/1/13}

\bibitem[{{Lindblad} {et~al.}(1996){Lindblad}, {Lindblad}, \&
  {Athanassoula}}]{1996A&A...313...65L}
{Lindblad}, P.~A.~B., {Lindblad}, P.~O., \& {Athanassoula}, E. 1996, \aap, 313,
  65

\bibitem[{{Lipman} {et~al.}(2024){Lipman}, {Battersby}, {Walker}, {Sormani},
  {Bally}, {Barnes}, {Ginsburg}, {Glover}, {Henshaw}, {Hatchfield}, {Immer},
  {Klessen}, {Longmore}, {Mills}, {Smith}, {Tress}, {Alboslani}, \&
  {Zhang}}]{2024arXiv241017321L}
{Lipman}, D., {Battersby}, C., {Walker}, D.~L., {et~al.} 2024, arXiv e-prints,
  arXiv:2410.17321, \dodoi{10.48550/arXiv.2410.17321}

\bibitem[{{Liszt}(2006)}]{2006A&A...447..533L}
{Liszt}, H.~S. 2006, \aap, 447, 533, \dodoi{10.1051/0004-6361:20054070}

\bibitem[{{Liszt}(2008)}]{2008A&A...486..467L}
---. 2008, \aap, 486, 467, \dodoi{10.1051/0004-6361:200809748}

\bibitem[{{{\L}okas}(2019)}]{2019A&A...629A..52L}
{{\L}okas}, E.~L. 2019, \aap, 629, A52, \dodoi{10.1051/0004-6361/201936056}

\bibitem[{{Lucey} {et~al.}(2023){Lucey}, {Pearson}, {Hunt}, {Hawkins}, {Ness},
  {Petersen}, {Price-Whelan}, \& {Weinberg}}]{2023MNRAS.520.4779L}
{Lucey}, M., {Pearson}, S., {Hunt}, J. A.~S., {et~al.} 2023, \mnras, 520, 4779,
  \dodoi{10.1093/mnras/stad406}

\bibitem[{{Maciejewski} {et~al.}(2002){Maciejewski}, {Teuben}, {Sparke}, \&
  {Stone}}]{2002MNRAS.329..502M}
{Maciejewski}, W., {Teuben}, P.~J., {Sparke}, L.~S., \& {Stone}, J.~M. 2002,
  \mnras, 329, 502, \dodoi{10.1046/j.1365-8711.2002.04957.x}

\bibitem[{{Mackey} {et~al.}(2024){Mackey}, {Morris}, {Ponti}, {Anastasopoulou},
  \& {Mondal}}]{2024ApJ...966L..32M}
{Mackey}, S.~C., {Morris}, M.~R., {Ponti}, G., {Anastasopoulou}, K., \&
  {Mondal}, S. 2024, \apjl, 966, L32, \dodoi{10.3847/2041-8213/ad3248}

\bibitem[{{Marshall} {et~al.}(2008){Marshall}, {Fux}, {Robin}, \&
  {Reyl{\'e}}}]{2008A&A...477L..21M}
{Marshall}, D.~J., {Fux}, R., {Robin}, A.~C., \& {Reyl{\'e}}, C. 2008, \aap,
  477, L21, \dodoi{10.1051/0004-6361:20078967}

\bibitem[{{Mart{\'\i}nez-Arranz} {et~al.}(2024){Mart{\'\i}nez-Arranz},
  {Sch{\"o}del}, {Nogueras-Lara}, {Najarro}, {Castellanos}, \&
  {Fedriani}}]{2024A&A...685L...7M}
{Mart{\'\i}nez-Arranz}, A., {Sch{\"o}del}, R., {Nogueras-Lara}, F., {et~al.}
  2024, \aap, 685, L7, \dodoi{10.1051/0004-6361/202449877}

\bibitem[{{Molinari} {et~al.}(2011){Molinari}, {Bally}, {Noriega-Crespo},
  {Compi{\`e}gne}, {Bernard}, {Paradis}, {Martin}, {Testi}, {Barlow}, {Moore},
  {Plume}, {Swinyard}, {Zavagno}, {Calzoletti}, {Di Giorgio}, {Elia},
  {Faustini}, {Natoli}, {Pestalozzi}, {Pezzuto}, {Piacentini}, {Polenta},
  {Polychroni}, {Schisano}, {Traficante}, {Veneziani}, {Battersby}, {Burton},
  {Carey}, {Fukui}, {Li}, {Lord}, {Morgan}, {Motte}, {Schuller},
  {Stringfellow}, {Tan}, {Thompson}, {Ward-Thompson}, {White}, \&
  {Umana}}]{2011ApJ...735L..33M}
{Molinari}, S., {Bally}, J., {Noriega-Crespo}, A., {et~al.} 2011, \apjl, 735,
  L33, \dodoi{10.1088/2041-8205/735/2/L33}

\bibitem[{{Nepal} {et~al.}(2024){Nepal}, {Chiappini}, {Guiglion}, {Steinmetz},
  {P{\'e}rez-Villegas}, {Queiroz}, {Miglio}, {Dohme}, \&
  {Khalatyan}}]{2024A&A...681L...8N}
{Nepal}, S., {Chiappini}, C., {Guiglion}, G., {et~al.} 2024, \aap, 681, L8,
  \dodoi{10.1051/0004-6361/202348365}

\bibitem[{{Nilipour} {et~al.}(2024){Nilipour}, {Ott}, {Meier}, {Svoboda},
  {Sormani}, {Ginsburg}, {Gramze}, {Butterfield}, \&
  {Klessen}}]{2024ApJ...977...37N}
{Nilipour}, A., {Ott}, J., {Meier}, D.~S., {et~al.} 2024, \apj, 977, 37,
  \dodoi{10.3847/1538-4357/ad8631}

\bibitem[{{Nogueras-Lara}(2024)}]{2024A&A...681L..21N}
{Nogueras-Lara}, F. 2024, \aap, 681, L21, \dodoi{10.1051/0004-6361/202348712}

\bibitem[{{Nonhebel} {et~al.}(2024){Nonhebel}, {Barnes}, {Immer},
  {Armijos-Abenda{\~n}o}, {Bally}, {Battersby}, {Burton}, {Butterfield},
  {Colzi}, {Garc{\'\i}a}, {Ginsburg}, {Henshaw}, {Hu}, {Jim{\'e}nez-Serra},
  {Klessen}, {Kruijssen}, {Liang}, {Longmore}, {Lu}, {Mart{\'\i}n}, {Mills},
  {Nogueras-Lara}, {Petkova}, {Pineda}, {Rivilla}, {S{\'a}nchez-Monge},
  {Santa-Maria}, {Smith}, {Sofue}, {Sormani}, {Tolls}, {Walker}, {Wallace},
  {Wang}, {Williams}, \& {Xu}}]{2024A&A...691A..70N}
{Nonhebel}, M., {Barnes}, A.~T., {Immer}, K., {et~al.} 2024, \aap, 691, A70,
  \dodoi{10.1051/0004-6361/202451190}

\bibitem[{{Papachristou} {et~al.}(2023){Papachristou}, {Dasyra},
  {Fern{\'a}ndez-Ontiveros}, {Audibert}, {Ruffa}, {Combes}, {Polkas}, \&
  {Gkogkou}}]{2023A&A...679A.115P}
{Papachristou}, M., {Dasyra}, K.~M., {Fern{\'a}ndez-Ontiveros}, J.~A., {et~al.}
  2023, \aap, 679, A115, \dodoi{10.1051/0004-6361/202346464}

\bibitem[{{Ponti} {et~al.}(2021){Ponti}, {Morris}, {Churazov}, {Heywood}, \&
  {Fender}}]{2021A&A...646A..66P}
{Ponti}, G., {Morris}, M.~R., {Churazov}, E., {Heywood}, I., \& {Fender}, R.~P.
  2021, \aap, 646, A66, \dodoi{10.1051/0004-6361/202039636}

\bibitem[{{Queiroz} {et~al.}(2021){Queiroz}, {Chiappini}, {Perez-Villegas},
  {Khalatyan}, {Anders}, {Barbuy}, {Santiago}, {Steinmetz}, {Cunha},
  {Schultheis}, {Majewski}, {Minchev}, {Minniti}, {Beaton}, {Cohen}, {da
  Costa}, {Fern{\'a}ndez-Trincado}, {Garcia-Hern{\'a}ndez}, {Geisler},
  {Hasselquist}, {Lane}, {Nitschelm}, {Rojas-Arriagada}, {Roman-Lopes},
  {Smith}, \& {Zasowski}}]{2021A&A...656A.156Q}
{Queiroz}, A.~B.~A., {Chiappini}, C., {Perez-Villegas}, A., {et~al.} 2021,
  \aap, 656, A156, \dodoi{10.1051/0004-6361/202039030}

\bibitem[{{Regan} \& {Teuben}(2004)}]{2004ApJ...600..595R}
{Regan}, M.~W., \& {Teuben}, P.~J. 2004, \apj, 600, 595, \dodoi{10.1086/380116}

\bibitem[{{Reid} {et~al.}(2016){Reid}, {Dame}, {Menten}, \&
  {Brunthaler}}]{2016ApJ...823...77R}
{Reid}, M.~J., {Dame}, T.~M., {Menten}, K.~M., \& {Brunthaler}, A. 2016, \apj,
  823, 77, \dodoi{10.3847/0004-637X/823/2/77}

\bibitem[{{Reid} {et~al.}(2019){Reid}, {Menten}, {Brunthaler}, {Zheng}, {Dame},
  {Xu}, {Li}, {Sakai}, {Wu}, {Immer}, {Zhang}, {Sanna}, {Moscadelli}, {Rygl},
  {Bartkiewicz}, {Hu}, {Quiroga-Nu{\~n}ez}, \& {van Langevelde}}]{Reid19}
{Reid}, M.~J., {Menten}, K.~M., {Brunthaler}, A., {et~al.} 2019, \apj, 885,
  131, \dodoi{10.3847/1538-4357/ab4a11}

\bibitem[{{Ridley} {et~al.}(2017){Ridley}, {Sormani}, {Tre{\ss}}, {Magorrian},
  \& {Klessen}}]{2017MNRAS.469.2251R}
{Ridley}, M. G.~L., {Sormani}, M.~C., {Tre{\ss}}, R.~G., {Magorrian}, J., \&
  {Klessen}, R.~S. 2017, \mnras, 469, 2251, \dodoi{10.1093/mnras/stx944}

\bibitem[{{Riener} {et~al.}(2019){Riener}, {Kainulainen}, {Henshaw}, {Orkisz},
  {Murray}, \& {Beuther}}]{2019A&A...628A..78R}
{Riener}, M., {Kainulainen}, J., {Henshaw}, J.~D., {et~al.} 2019, \aap, 628,
  A78, \dodoi{10.1051/0004-6361/201935519}

\bibitem[{{Rodriguez-Fernandez} \& {Combes}(2008)}]{2008A&A...489..115R}
{Rodriguez-Fernandez}, N.~J., \& {Combes}, F. 2008, \aap, 489, 115,
  \dodoi{10.1051/0004-6361:200809644}

\bibitem[{{Rodriguez-Fernandez} {et~al.}(2006){Rodriguez-Fernandez}, {Combes},
  {Martin-Pintado}, {Wilson}, \& {Apponi}}]{2006A&A...455..963R}
{Rodriguez-Fernandez}, N.~J., {Combes}, F., {Martin-Pintado}, J., {Wilson},
  T.~L., \& {Apponi}, A. 2006, \aap, 455, 963,
  \dodoi{10.1051/0004-6361:20064813}

\bibitem[{{Ruffa} {et~al.}(2020){Ruffa}, {Laing}, {Prandoni}, {Paladino},
  {Parma}, {Davis}, \& {Bureau}}]{2020MNRAS.499.5719R}
{Ruffa}, I., {Laing}, R.~A., {Prandoni}, I., {et~al.} 2020, \mnras, 499, 5719,
  \dodoi{10.1093/mnras/staa3166}

\bibitem[{{Ruffa} {et~al.}(2019){Ruffa}, {Davis}, {Prandoni}, {Laing},
  {Paladino}, {Parma}, {de Ruiter}, {Casasola}, {Bureau}, \&
  {Warren}}]{2019MNRAS.489.3739R}
{Ruffa}, I., {Davis}, T.~A., {Prandoni}, I., {et~al.} 2019, \mnras, 489, 3739,
  \dodoi{10.1093/mnras/stz2368}

\bibitem[{{Sarkar}(2024)}]{2024A&ARv..32....1S}
{Sarkar}, K.~C. 2024, \aapr, 32, 1, \dodoi{10.1007/s00159-024-00152-1}

\bibitem[{{Schinnerer} \& {Leroy}(2024)}]{2024ARA&A..62..369S}
{Schinnerer}, E., \& {Leroy}, A.~K. 2024, \araa, 62, 369,
  \dodoi{10.1146/annurev-astro-071221-052651}

\bibitem[{{Schinnerer} {et~al.}(2023){Schinnerer}, {Emsellem}, {Henshaw},
  {Liu}, {Meidt}, {Querejeta}, {Renaud}, {Sormani}, {Sun}, {Egorov}, {Larson},
  {Leroy}, {Rosolowsky}, {Sandstrom}, {Williams}, {Barnes}, {Bigiel},
  {Chevance}, {Cao}, {Chandar}, {Dale}, {Eibensteiner}, {Glover}, {Grasha},
  {Hannon}, {Hassani}, {Kim}, {Klessen}, {Kruijssen}, {Murphy}, {Neumann},
  {Pan}, {Pety}, {Saito}, {Stuber}, {Tre{\ss}}, {Usero}, {Watkins}, {Whitmore},
  \& {Phangs}}]{2023ApJ...944L..15S}
{Schinnerer}, E., {Emsellem}, E., {Henshaw}, J.~D., {et~al.} 2023, \apjl, 944,
  L15, \dodoi{10.3847/2041-8213/acac9e}

\bibitem[{{Schuller} {et~al.}(2021){Schuller}, {Urquhart}, {Csengeri},
  {Colombo}, {Duarte-Cabral}, {Mattern}, {Ginsburg}, {Pettitt}, {Wyrowski},
  {Anderson}, {Azagra}, {Barnes}, {Beltran}, {Beuther}, {Billington},
  {Bronfman}, {Cesaroni}, {Dobbs}, {Eden}, {Lee}, {Medina}, {Menten}, {Moore},
  {Montenegro-Montes}, {Ragan}, {Rigby}, {Riener}, {Russeil}, {Schisano},
  {Sanchez-Monge}, {Traficante}, {Zavagno}, {Agurto}, {Bontemps}, {Finger},
  {Giannetti}, {Gonzalez}, {Hernandez}, {Henning}, {Kainulainen}, {Kauffmann},
  {Leurini}, {Lopez}, {Mac-Auliffe}, {Mazumdar}, {Molinari}, {Motte}, {Muller},
  {Nguyen-Luong}, {Parra}, {Perez-Beaupuits}, {Schilke}, {Schneider}, {Suri},
  {Testi}, {Torstensson}, {Veena}, {Venegas}, {Wang}, \&
  {Wienen}}]{2021MNRAS.500.3064S}
{Schuller}, F., {Urquhart}, J.~S., {Csengeri}, T., {et~al.} 2021, \mnras, 500,
  3064, \dodoi{10.1093/mnras/staa2369}

\bibitem[{{Schultheis} {et~al.}(2014){Schultheis}, {Chen}, {Jiang}, {Gonzalez},
  {Enokiya}, {Fukui}, {Torii}, {Rejkuba}, \& {Minniti}}]{2014A&A...566A.120S}
{Schultheis}, M., {Chen}, B.~Q., {Jiang}, B.~W., {et~al.} 2014, \aap, 566,
  A120, \dodoi{10.1051/0004-6361/201322788}

\bibitem[{{Shen} \& {Zheng}(2020)}]{2020RAA....20..159S}
{Shen}, J., \& {Zheng}, X.-W. 2020, Research in Astronomy and Astrophysics, 20,
  159, \dodoi{10.1088/1674-4527/20/10/159}

\bibitem[{{Sheth} {et~al.}(2005){Sheth}, {Vogel}, {Regan}, {Thornley}, \&
  {Teuben}}]{2005ApJ...632..217S}
{Sheth}, K., {Vogel}, S.~N., {Regan}, M.~W., {Thornley}, M.~D., \& {Teuben},
  P.~J. 2005, \apj, 632, 217, \dodoi{10.1086/432409}

\bibitem[{{Simion} {et~al.}(2017){Simion}, {Belokurov}, {Irwin}, {Koposov},
  {Gonzalez-Fernandez}, {Robin}, {Shen}, \& {Li}}]{2017MNRAS.471.4323S}
{Simion}, I.~T., {Belokurov}, V., {Irwin}, M., {et~al.} 2017, \mnras, 471,
  4323, \dodoi{10.1093/mnras/stx1832}

\bibitem[{{Sormani} \& {Barnes}(2019)}]{2019MNRAS.484.1213S}
{Sormani}, M.~C., \& {Barnes}, A.~T. 2019, \mnras, 484, 1213,
  \dodoi{10.1093/mnras/stz046}

\bibitem[{{Sormani} {et~al.}(2015){Sormani}, {Binney}, \&
  {Magorrian}}]{2015MNRAS.449.2421S}
{Sormani}, M.~C., {Binney}, J., \& {Magorrian}, J. 2015, \mnras, 449, 2421,
  \dodoi{10.1093/mnras/stv441}

\bibitem[{{Sormani} {et~al.}(2024){Sormani}, {Sobacchi}, \&
  {Sanders}}]{2024MNRAS.528.5742S}
{Sormani}, M.~C., {Sobacchi}, E., \& {Sanders}, J.~L. 2024, \mnras, 528, 5742,
  \dodoi{10.1093/mnras/stae082}

\bibitem[{{Sormani} {et~al.}(2020){Sormani}, {Tress}, {Glover}, {Klessen},
  {Battersby}, {Clark}, {Hatchfield}, \& {Smith}}]{2020MNRAS.497.5024S}
{Sormani}, M.~C., {Tress}, R.~G., {Glover}, S. C.~O., {et~al.} 2020, \mnras,
  497, 5024, \dodoi{10.1093/mnras/staa1999}

\bibitem[{{Sormani} {et~al.}(2019){Sormani}, {Tre{\ss}}, {Glover}, {Klessen},
  {Barnes}, {Battersby}, {Clark}, {Hatchfield}, \&
  {Smith}}]{2019MNRAS.488.4663S}
{Sormani}, M.~C., {Tre{\ss}}, R.~G., {Glover}, S. C.~O., {et~al.} 2019, \mnras,
  488, 4663, \dodoi{10.1093/mnras/stz2054}

\bibitem[{{Sormani} {et~al.}(2022){Sormani}, {Sanders}, {Fritz}, {Smith},
  {Gerhard}, {Sch{\"o}del}, {Magorrian}, {Neumayer}, {Nogueras-Lara},
  {Feldmeier-Krause}, {Mastrobuono-Battisti}, {Schultheis}, {Shahzamanian},
  {Vasiliev}, {Klessen}, {Lucas}, \& {Minniti}}]{2022MNRAS.512.1857S}
{Sormani}, M.~C., {Sanders}, J.~L., {Fritz}, T.~K., {et~al.} 2022, \mnras, 512,
  1857, \dodoi{10.1093/mnras/stac639}

\bibitem[{{Sormani} {et~al.}(2023){Sormani}, {Barnes}, {Sun}, {Stuber},
  {Schinnerer}, {Emsellem}, {Leroy}, {Glover}, {Henshaw}, {Meidt}, {Neumann},
  {Querejeta}, {Williams}, {Bigiel}, {Eibensteiner}, {Fragkoudi}, {Levy},
  {Grasha}, {Klessen}, {Kruijssen}, {Neumayer}, {Pinna}, {Rosolowsky}, {Smith},
  {Teng}, {Tress}, \& {Watkins}}]{2023MNRAS.523.2918S}
{Sormani}, M.~C., {Barnes}, A.~T., {Sun}, J., {et~al.} 2023, \mnras, 523, 2918,
  \dodoi{10.1093/mnras/stad1554}

\bibitem[{{Stark} \& {Bania}(1986)}]{1986ApJ...306L..17S}
{Stark}, A.~A., \& {Bania}, T.~M. 1986, \apjl, 306, L17, \dodoi{10.1086/184695}

\bibitem[{{Strong} {et~al.}(2004){Strong}, {Moskalenko}, {Reimer}, {Digel}, \&
  {Diehl}}]{2004A&A...422L..47S}
{Strong}, A.~W., {Moskalenko}, I.~V., {Reimer}, O., {Digel}, S., \& {Diehl}, R.
  2004, \aap, 422, L47, \dodoi{10.1051/0004-6361:20040172}

\bibitem[{{Stuber} {et~al.}(2023){Stuber}, {Schinnerer}, {Williams},
  {Querejeta}, {Meidt}, {Emsellem}, {Barnes}, {Klessen}, {Leroy}, {Neumann},
  {Sormani}, {Bigiel}, {Chevance}, {Dale}, {Faesi}, {Glover}, {Grasha},
  {Kruijssen}, {Liu}, {Pan}, {Pety}, {Pinna}, {Saito}, {Usero}, \&
  {Watkins}}]{2023A&A...676A.113S}
{Stuber}, S.~K., {Schinnerer}, E., {Williams}, T.~G., {et~al.} 2023, \aap, 676,
  A113, \dodoi{10.1051/0004-6361/202346318}

\bibitem[{{Su} {et~al.}(2016){Su}, {Sun}, {Li}, {Zhang}, {Zhou}, {Fang},
  {Yang}, \& {Chen}}]{2016ApJ...828...59S}
{Su}, Y., {Sun}, Y., {Li}, C., {et~al.} 2016, \apj, 828, 59,
  \dodoi{10.3847/0004-637X/828/1/59}

\bibitem[{{Su} {et~al.}(2019){Su}, {Yang}, {Zhang}, {Gong}, {Wang}, {Zhou},
  {Wang}, {Chen}, {Sun}, {Chen}, {Xu}, \& {Jiang}}]{2019ApJS..240....9S}
{Su}, Y., {Yang}, J., {Zhang}, S., {et~al.} 2019, \apjs, 240, 9,
  \dodoi{10.3847/1538-4365/aaf1c8}

\bibitem[{{Su} {et~al.}(2021){Su}, {Yang}, {Yan}, {Zhang}, {Wang}, {Sun},
  {Chen}, {Wang}, {Zhou}, {Chen}, {Jiang}, \& {Wang}}]{2021ApJ...910..131S}
{Su}, Y., {Yang}, J., {Yan}, Q.-Z., {et~al.} 2021, \apj, 910, 131,
  \dodoi{10.3847/1538-4357/abe5ab}

\bibitem[{{Su} {et~al.}(2022){Su}, {Zhang}, {Yang}, {Yan}, {Sun}, {Wang},
  {Zhang}, {Chen}, {Chen}, {Zhou}, \& {Yuan}}]{2022ApJ...930..112S}
{Su}, Y., {Zhang}, S., {Yang}, J., {et~al.} 2022, \apj, 930, 112,
  \dodoi{10.3847/1538-4357/ac63b3}

\bibitem[{{Su} {et~al.}(2024){Su}, {Zhang}, {Sun}, {Yang}, {Yan}, {Zhang},
  {Chen}, {Chen}, {Zhou}, \& {Yuan}}]{2024ApJ...971L...6S}
{Su}, Y., {Zhang}, S., {Sun}, Y., {et~al.} 2024, \apjl, 971, L6,
  \dodoi{10.3847/2041-8213/ad656d}

\bibitem[{{Takekawa} {et~al.}(2024){Takekawa}, {Oka}, {Tsujimoto}, {Yokozuka},
  {Harada}, {Kaneko}, {Enokiya}, \& {Iwata}}]{2024ApJ...972L...3T}
{Takekawa}, S., {Oka}, T., {Tsujimoto}, S., {et~al.} 2024, \apjl, 972, L3,
  \dodoi{10.3847/2041-8213/ad6c51}

\bibitem[{{Tokuyama} {et~al.}(2019){Tokuyama}, {Oka}, {Takekawa}, {Iwata},
  {Tsujimoto}, {Yamada}, {Furusawa}, \& {Nomura}}]{2019PASJ...71S..19T}
{Tokuyama}, S., {Oka}, T., {Takekawa}, S., {et~al.} 2019, \pasj, 71, S19,
  \dodoi{10.1093/pasj/psy150}

\bibitem[{{Tress} {et~al.}(2020){Tress}, {Sormani}, {Glover}, {Klessen},
  {Battersby}, {Clark}, {Hatchfield}, \& {Smith}}]{2020MNRAS.499.4455T}
{Tress}, R.~G., {Sormani}, M.~C., {Glover}, S. C.~O., {et~al.} 2020, \mnras,
  499, 4455, \dodoi{10.1093/mnras/staa3120}

\bibitem[{{Tress} {et~al.}(2024){Tress}, {Sormani}, {Girichidis}, {Glover},
  {Klessen}, {Smith}, {Sobacchi}, {Armillotta}, {Barnes}, {Battersby}, {Bogue},
  {Brucy}, {Colzi}, {Federrath}, {Garc{\'\i}a}, {Ginsburg}, {G{\"o}ller},
  {Hatchfield}, {Henkel}, {Hennebelle}, {Henshaw}, {Hirschmann}, {Hu},
  {Kauffmann}, {Kruijssen}, {Lazarian}, {Lipman}, {Longmore}, {Morris},
  {Nogueras-Lara}, {Petkova}, {Pillai}, {Rivilla}, {S{\'a}nchez-Monge},
  {Soler}, {Whitworth}, \& {Zhang}}]{2024A&A...691A.303T}
{Tress}, R.~G., {Sormani}, M.~C., {Girichidis}, P., {et~al.} 2024, \aap, 691,
  A303, \dodoi{10.1051/0004-6361/202450035}

\bibitem[{{Veena} {et~al.}(2024){Veena}, {Kim}, {S{\'a}nchez-Monge}, {Schilke},
  {Menten}, {Fuller}, {Sormani}, {Wyrowski}, {Banda-Barrag{\'a}n}, {Riquelme},
  {Tarr{\'\i}o}, \& {de Vicente}}]{2024A&A...689A.121V}
{Veena}, V.~S., {Kim}, W.~J., {S{\'a}nchez-Monge}, {\'A}., {et~al.} 2024, \aap,
  689, A121, \dodoi{10.1051/0004-6361/202450902}

\bibitem[{{Vislosky} {et~al.}(2024){Vislosky}, {Minchev}, {Khoperskov},
  {Martig}, {Buck}, {Hilmi}, {Ratcliffe}, {Bland-Hawthorn}, {Quillen},
  {Steinmetz}, \& {de Jong}}]{2024MNRAS.528.3576V}
{Vislosky}, E., {Minchev}, I., {Khoperskov}, S., {et~al.} 2024, \mnras, 528,
  3576, \dodoi{10.1093/mnras/stae083}

\bibitem[{{Walker} {et~al.}(2024){Walker}, {Battersby}, {Lipman}, {Sormani},
  {Ginsburg}, {Glover}, {Henshaw}, {Longmore}, {Klessen}, {Immer}, {Alboslani},
  {Bally}, {Barnes}, {Hatchfield}, {Mills}, {Smith}, {Tress}, \&
  {Zhang}}]{2024arXiv241017320W}
{Walker}, D.~L., {Battersby}, C., {Lipman}, D., {et~al.} 2024, arXiv e-prints,
  arXiv:2410.17320, \dodoi{10.48550/arXiv.2410.17320}

\bibitem[{{Wallace} {et~al.}(2022){Wallace}, {Battersby}, {Mills}, {Henshaw},
  {Sormani}, {Ginsburg}, {Barnes}, {Hatchfield}, {Glover}, \&
  {Anderson}}]{2022ApJ...939...58W}
{Wallace}, J., {Battersby}, C., {Mills}, E.~A.~C., {et~al.} 2022, \apj, 939,
  58, \dodoi{10.3847/1538-4357/ac951a}

\bibitem[{{Wang} {et~al.}(2023){Wang}, {Feng}, {Yang}, {Chen}, {Su}, {Yan},
  {Du}, {Ma}, \& {Cai}}]{2023AJ....166..121W}
{Wang}, C., {Feng}, H., {Yang}, J., {et~al.} 2023, \aj, 166, 121,
  \dodoi{10.3847/1538-3881/acebdd}

\bibitem[{{Wegg} {et~al.}(2015){Wegg}, {Gerhard}, \&
  {Portail}}]{2015MNRAS.450.4050W}
{Wegg}, C., {Gerhard}, O., \& {Portail}, M. 2015, \mnras, 450, 4050,
  \dodoi{10.1093/mnras/stv745}

\bibitem[{{Wheeler} {et~al.}(2022){Wheeler}, {Abril-Cabezas}, {Trick},
  {Fragkoudi}, \& {Ness}}]{2022ApJ...935...28W}
{Wheeler}, A., {Abril-Cabezas}, I., {Trick}, W.~H., {Fragkoudi}, F., \& {Ness},
  M. 2022, \apj, 935, 28, \dodoi{10.3847/1538-4357/ac7da0}

\bibitem[{{Whitmore} {et~al.}(2023){Whitmore}, {Chandar}, {Rodr{\'\i}guez},
  {Lee}, {Emsellem}, {Floyd}, {Kim}, {Kruijssen}, {Mok}, {Sormani}, {Boquien},
  {Dale}, {Faesi}, {Henny}, {Hannon}, {Thilker}, {White}, {Barnes}, {Bigiel},
  {Chevance}, {Henshaw}, {Klessen}, {Leroy}, {Liu}, {Maschmann}, {Meidt},
  {Rosolowsky}, {Schinnerer}, {Sun}, {Watkins}, \&
  {Williams}}]{2023ApJ...944L..14W}
{Whitmore}, B.~C., {Chandar}, R., {Rodr{\'\i}guez}, M.~J., {et~al.} 2023,
  \apjl, 944, L14, \dodoi{10.3847/2041-8213/acae94}

\bibitem[{{Williams} {et~al.}(2022){Williams}, {Walker}, {Longmore}, {Barnes},
  {Battersby}, {Garay}, {Ginsburg}, {Gomez}, {Henshaw}, {Ho}, {Kruijssen},
  {Lu}, {Mills}, {Petkova}, \& {Zhang}}]{2022MNRAS.514..578W}
{Williams}, B.~A., {Walker}, D.~L., {Longmore}, S.~N., {et~al.} 2022, \mnras,
  514, 578, \dodoi{10.1093/mnras/stac1378}

\bibitem[{{Wylie} {et~al.}(2022){Wylie}, {Clarke}, \&
  {Gerhard}}]{2022A&A...659A..80W}
{Wylie}, S.~M., {Clarke}, J.~P., \& {Gerhard}, O.~E. 2022, \aap, 659, A80,
  \dodoi{10.1051/0004-6361/202142343}

\bibitem[{{Zhang} {et~al.}(2024{\natexlab{a}}){Zhang}, {Belokurov}, {Evans},
  {Kane}, \& {Sanders}}]{2024MNRAS.533.3395Z}
{Zhang}, H., {Belokurov}, V., {Evans}, N.~W., {Kane}, S.~G., \& {Sanders},
  J.~L. 2024{\natexlab{a}}, \mnras, 533, 3395, \dodoi{10.1093/mnras/stae2023}

\bibitem[{{Zhang} {et~al.}(2024{\natexlab{b}}){Zhang}, {Su}, {Chen}, {Fang},
  {Yan}, {Zhang}, {Sun}, {Wang}, {Feng}, {Ma}, {Zhang}, {Zhuang}, {Zhou},
  {Chen}, \& {Yang}}]{2024AJ....167..220Z}
{Zhang}, S., {Su}, Y., {Chen}, X., {et~al.} 2024{\natexlab{b}}, \aj, 167, 220,
  \dodoi{10.3847/1538-3881/ad2fcb}

\end{thebibliography}
\bibliographystyle{aasjournal}

\begin{deluxetable}{ccccccccccc}
\tabletypesize{\scriptsize}
\tablecaption{Parameters of MC Samples toward the CMZ}
\tablehead{
\colhead{\begin{tabular}{c}
Name  \\
\\
\end{tabular}} &
\colhead{\begin{tabular}{c}
Number$^{\mathit {a}}$            \\
     \\
\end{tabular}} &
\colhead{\begin{tabular}{c}
$T_{\rm {peak12}}$$^{\mathit {b}}$     \\
(K)          \\
\end{tabular}} &
\colhead{\begin{tabular}{c}
$T_{\rm {peak13}}$$^{\mathit {b}}$     \\
(K)          \\
\end{tabular}} &
\colhead{\begin{tabular}{c}
$v({\rm {FWHM})}$$^{\mathit {b}}$      \\
(km~s$^{-1}$)          \\
\end{tabular}} &
\colhead{\begin{tabular}{c}
$\frac{I_{^{13}{\rm CO}}}{I_{^{12}{\rm CO}}}^{\mathit {b}}$             \\
      \\
\end{tabular}} &
\colhead{\begin{tabular}{c}
$X_{\rm CO}$$^{\mathit {c}}$     \\
($\E{20}$~cm$^{-2}$(K~km~s$^{-1})^{-1}$)            \\
\end{tabular}} &
\colhead{\begin{tabular}{c}
MC Mass$^{\mathit {d}}$   \\
($\E{6}\Msun$)              \\
\end{tabular}} &
\colhead{\begin{tabular}{c}
Total Gas Mass   \\
($\E{6}\Msun$)              \\
\end{tabular}}
}
\startdata
DL1                           &  64  & 8.7$\pm$3.2  & 3.5$\pm$1.8  & 15.5$\pm$4.3 &
0.14$\pm$0.04  & 0.9$\pm$0.4 &  5.6$\pm2.4$   & $\sim9.0\pm3.3$$^{\mathit {e}}$  \\
Intermediate$^{\mathit {f}}$ &  93 & 9.4$\pm$3.7  & 4.3$\pm$2.2  & 18.1$\pm$6.0 &
$\lsim$0.17$\pm$0.04$^{\mathit {g}}$ & 1.2$\pm$0.5  &  10.0$\pm4.3$  & $\gsim17.0\pm7.3$$^{\mathit {h}}$ \\
Overshoot                     &  13  & 5.5$\pm$2.0  & 1.6$\pm$0.3  & 10.3$\pm$3.2 &
0.10$\pm$0.02  & 0.7$\pm$0.3  &  0.2$\pm0.1$   & $\gsim0.9\pm0.3$$^{\mathit {i}}$   \\
\hline
\enddata
\tablecomments{
$^{\mathit {a}}$ Here we only list the numbers of the large-scale MC samples with
an angular size $>50 \ {\rm {arcmin}}^2$.  \\
$^{\mathit {b}}$ Statistics from \twCO\ flux-weighted MC samples. \\
$^{\mathit {c}}$ $X_{\rm CO}$ is derived from the \twCO\ and \thCO\ emission of the
large-scale MCs \citep[i.e., $N({{\rm {H}}_2})\approx7\pm3\E{5}\times N({^{13}{\rm {CO}}})$;
see the details in][]{2024ApJ...971L...6S}. \\
$^{\mathit {d}}$ The total mass of the large-scale MCs is calculated from the
estimated CO-to-H$_2$ conversion factor.  \\
$^{\mathit {e}}$ The total gas mass is calculated from the CO-to-H$_2$ conversion
factor for all identified MCs along the near dust lane. The contribution from
the corresponding \HI\ gas, accounting for $\sim$15\% of the molecular gas mass,
has been included in the calculations of the total gas mass.   \\
$^{\mathit {f}}$ MCs with $v_{\rm {LSR}}$ between the DL1 and the overshoot gas
(Figure~\ref{fig:f1}). Note that some MC samples from the blended regions have been
excluded, so the reported MC mass is a lower limit. These MCs may also include a small
portion of the fresh gas in DL1 due to interactions between gas flows in the bar channel.\\
$^{\mathit {g}}$ The total emission of \thCO\ is calculated from the $l$-$b$-$v$ 
space of the fitted \twCO\ cloud. The measurement of $I_{^{13}{\rm CO}}$ can be 
contaminated with unrelated gas emission due to the complicated velocity components
toward the large MCs. Consequently, the ratio reported here is only an upper limit.\\
$^{\mathit {h}}$ Because of the confusion from foreground MCs near the Galactic plane,
only some of small-scale MCs associated with the shock regions, such as G3, G5,
DL2, and DL3, are considered here and are conservatively estimated to
account for $\gsim70\%$ of the total mass of the large-scale MCs. \\
$^{\mathit {i}}$ The small MCs dominate the mass in the region, and the contribution
of the corresponding \HI\ gas accounts for $\gsim50\%$ of the total molecular gas mass.
}
\end{deluxetable}

\clearpage

\clearpage

\clearpage

\end{document}